\documentclass[preprint, 5p, number, times, sort&compress]{elsarticle} 

\usepackage{amssymb}

\usepackage{hyperref} 
\usepackage{algorithm}
\usepackage{algpseudocode}
\usepackage{amsmath} 

\usepackage{multirow}
\usepackage{tabularx}
\newcolumntype{C}{>{\centering\arraybackslash}X}
\newcolumntype{D}{>{\hsize=\dimexpr2\hsize+2\tabcolsep+\arrayrulewidth\relax}C}

\usepackage{xr}
\usepackage{color, soul}

\begin{document}

\begin{frontmatter}



\title{Using Bursty Announcements for Detecting BGP Routing Anomalies}


\author[ornl]{Pablo~Moriano\corref{cor1}}
\ead{moriano@ornl.gov}

\author[spelman]{Raquel~Hill}
\ead{raquel.hill@spelman.edu}

\author[iu]{L.~Jean~Camp}
\ead{ljcamp@indiana.edu}

\cortext[cor1]{Corresponding author. This manuscript has been authored by UT-Battelle, LLC under Contract No. DE-AC05-00OR22725 with the U.S. Department of Energy. The United States Government retains and the publisher, by accepting the article for publication, acknowledges that the United States Government retains a non-exclusive, paid-up, irrevocable, worldwide license to publish or reproduce the published form of this manuscript, or allow others to do so, for United States Government purposes. The Department of Energy will provide public access to these results of federally sponsored research in accordance with the DOE Public Access Plan (http://energy.gov/downloads/doe-public-access-plan).}
\address[ornl]{Computer Science and Mathematics Division, Oak Ridge National Laboratory, Oak Ridge, TN  37830, USA}
\address[spelman]{Computer and Information Sciences Department, Spelman College, Atlanta, GA 30314, USA}
\address[iu]{Luddy School of Informatics, Computing, and Engineering, Indiana University, Bloomington, IN 47408, USA}

\address{}

\begin{abstract}

Despite the robust structure of the Internet, it is still susceptible to disruptive routing updates that prevent network traffic from reaching its destination. Our research shows that BGP announcements that are associated with disruptive updates tend to occur in groups of relatively high frequency, followed by periods of infrequent activity. We hypothesize that we may use these bursty characteristics to detect anomalous routing incidents. In this work, we use manually verified ground truth metadata and volume of announcements as a baseline measure, and propose a burstiness measure that detects prior anomalous incidents with high recall and better precision than the volume baseline. We quantify the burstiness of inter-arrival times around the date and times of four large-scale incidents: the Indosat hijacking event in April 2014, the Telecom Malaysia leak in June 2015, the Bharti Airtel Ltd. hijack in November 2015, and the MainOne leak in November 2018; and three smaller scale incidents that led to traffic interception: the Belarusian traffic direction in February 2013, the Icelandic traffic direction in July 2013, and the Russian telecom that hijacked financial services in April 2017. Our method leverages the burstiness of disruptive update messages to detect these incidents. We describe limitations, open challenges, and how this method can be used for routing anomaly detection.

\end{abstract}

%

\begin{keyword}
Border Gateway Protocol (BGP) \sep Internet measurements \sep prefix hijacking \sep time series analysis \sep anomaly detection \sep burstiness



\end{keyword}

\end{frontmatter}



\section{Introduction} \label{Introduction}

The Internet, although extremely robust \cite{doyle2005robust}, is notoriously vulnerable to attack by means of the Border Gateway Protocol (BGP) \cite{butler2010survey}. BGP update messages are assumed to be trustworthy. In other words, the reachability information shared between autonomous systems (ASes) is assumed to be correct without any verification. Despite the fact that the latest version of the BGP protocol was released in 2006 \cite{rfc4271}, there are no inherent protection mechanisms against participants advertising false routes. 

In practice, BGP lacks authentication mechanisms not only for the announcement of the origin of IP prefixes but also the paths to that prefix. This leaves BGP vulnerable to unintended misconfiguration and malicious attacks \cite{Goldberg:2014:Taking:So:Long:Secure:Internet}. The results of these disruptions include traffic blackholing and traffic interception. In traffic blackholing, the network traffic is dropped, never reaching its destination \cite{Brown:2008youtube}. In traffic interception, the announcing AS reroutes traffic for the victim IP prefix and redirects it to the original origin AS after interception \cite{Cowie:2013belarus}. On this misdirected route, the traffic may be subject to eavesdropping \cite{arnbak2015loopholes}, traffic analysis \cite{sun2015raptor}, or tampering \cite{Shaw:SpamhausHijack}. 

Well-known examples of BGP anomalies include the 2014 incident in which Indonesia's largest communication provider (Indosat) \cite{Toonk:Indonesia-Hijack, Zmijewski:IndonesiaHijack} hijacked more than 320,000 routes, or roughly two-thirds of the Internet, for almost three hours, as well as the 2017 event in which Rostelecom, one of Russia's largest, partially state-owned Internet service providers \cite{Goodin:2017:Russian:BGP:MITM:Attack:ARS:Technica, Toonk:2017:Russian:BGP:MITM:Cisco} hijacked not only prefixes belonging to financial services firms but also e-commerce and payment services. Both were identified only after widespread diffusion of the incorrect routing information. In the case of Rostelecom, it was confirmed that the traffic passed through Rostelecom en route to its intended destinations.

The current approaches to the challenges of routing anomalies rely on $(i)$ cryptographic authentication or $(ii)$ anomaly detection. Cryptographic protocols include the Resource Public Key Infrastructure (RPKI) \cite{rfc6480} for origin authentication and BGPsec \cite{RFC18BGPSec}, which offers the ability to authenticate an entire path. These approaches are powerful, but there has not been widespread adoption \cite{Gill:2011:TansitiontoRPKI}. This lack of adoption may be caused by processor requirements, memory requirements, or a lack of incentive alignment \cite{Hall:2014:Collaborating:With:Enemy}. BGPsec also is not widely used because it does not support partial deployment, and cryptographic solutions are expensive. Perhaps more importantly, it has been shown that even with their widespread adoption, it will be not possible to avoid the occurrence of route leaks, such as the Telecom Malaysia incident in 2015 \cite{Toonk:TelecomMalasya-Leak, Madory:2015:Dyn:Malaysia:Leak}.

Anomaly detection approaches rely on measuring the control-plane (using BGP feeds) or the data-plane (exploring reachability of IP addresses in suspicious announced routes), or a combination of both. Anomaly detection does not require changes in the protocol itself. They primarily are used in detecting anomalies based on passive or active measurements in order to alert operators for mitigation and response  \cite{Khare:2012MOAS, Zhang:2010:iSPY, Shi:2012:Argus, Toonk:BGPmon}. Most anomaly detection approaches are reactive because they identify harm after disruptive updates have polluted some detectable threshold of ASes with fake announcements.

Here, we propose an anomaly detection method that aims to identify incipient incidents before diffusion and harm by identifying  a routing event as it emerges. Our goal is to identify anomalous routing events as soon as possible with respect to manually verified ground truth metadata, such as BGPMon~\cite{BGPMon:Blog}, Oracle Dyn~\cite{Dyn:2019:Security:Blog}, and Ars Technica~\cite{Ars-Technica:IT:Blog}. Our ground truth sources manually investigate these incidents and determine its impact based on empirical measurements. A similar procedure has been used before for labeling purposes~\cite{Cho:2019:BGP:Hijacking:Classification}. 
To do this, we use control-plane data collected by RouteViews \cite{Meyer:RouteViews} and served by BGPStream \cite{Orsini:2016:BGPStream}. The key observation in our anomaly detection method is that there are bursty BGP announcements as soon as new routes are adopted by neighbor ASes. We characterize bursty announcements through statistical analysis of inter-arrival times. We conduct a case-based systematic analysis of the changes of inter-arrival times that are associated with well-known anomalous events that resulted in traffic blackholing and intersection. We then propose a method based on control-plane information and statistical analysis to detect anomalous BGP announcements. We show that the proposed method is able to correctly identify the incidents in agreement with the ground truth while reducing significantly the number of false positives when compared with the volume baseline. 

\subsection{Our Contributions} \label{Our Contributions}

In this paper, we make the following contributions.

First, we validate our conjecture that inter-arrival time patterns of BGP announcements are a useful signature for identification of BGP routing incidents. We show that bursty patterns of announcements are noticeable in agreement with the manually verified ground truth metadata. 
To do so, we quantify the burstiness of BGP announcements by observing that when there are BGP incidents, there are groups of announcements with short inter-arrival times followed by larger ones. We report that this observation is independent of the volume of announcements. 

Second, we describe the design of a proof-of-concept BGP anomaly detection method that uses data only from current route collectors. We use RouteViews route collectors to compute a detection signature of BGP incidents based on the impact of short inter-arrival times. 

Third, we report results of a longitudinal analysis of large-scale routing incidents. We evaluated the proposed method by studying four different BGP incidents that resulted in blackholing, i.e., Indosat in April 2014, Telecom Malaysia in June 2015, Bharti Airtel Ltd. in November 2015, and MainOne in November 2018; and three additional incidents that resulted in traffic interception, i.e., GlobalOneBell in February 2013, Opin Kerfi in July 2013, and Rostelecom in April 2017. Our approach allows for statistically significant differentiation between normal behavior and disruption or anomalous changes during the incidents. We validate this by conducting Monte Carlo simulations on the burstiness behavior of ASes. We show that the proposed method outperforms, in most cases, the performance achieved by the baseline of volume of announcements in terms of false positives and negatives. 

We leverage a better understanding of past incidents to deter future incidents in the network. We hope that this study will inspire more investigations that maximize the use of routing updates, at the collector level, for BGP hijacking identification. We provide access to the data collection and analysis scripts for reproducibility purposes.\footnote{Auxiliary material can be found at \url{https://github.com/pmoriano/bgp-burstiness}.} 

\subsection{Related Work} \label{Related work}

\subsubsection{Prior Works Closely Related to the Present Study}

Zhang \emph{et al.} \cite{Zhang:2004:Anomalous:Routing:Dynamics} developed signature- and statistic-based approaches to detect anomalous BGP routing dynamics. Their statistic-based approach used five measures to model BGP updates, including an intensity measure derived from update inter-arrival times. They discussed advantages and disadvantages from their two approaches suggesting a combination of them to achieve better performance.  


Chen \emph{et al.} \cite{Chen:2016:Large:Scale:BGP:Detection} proposed a statistical method for detecting large-scale high impact BGP incidents relying on the update visibility matrix, i.e., a binary matrix in which data from each collector and prefix is recorded. The core of their contribution is on proposing a heuristic algorithm that finds a submatrix that is dense and large from the original matrix (which indicates unusual activity seen from different collectors affecting similar prefixes). They validated their results by showing that the identified events are strongly correlated with well-known large-scale incidents.

Zhang \emph{et al.} \cite{Zhang:2017:Seismograph} proposed I-seismograph, a two-phase clustering method to discover abnormal routing attributes extracted from BGP updates \cite{li:2007:BGP:Dynamics}. Their anomaly detection method estimates normal dynamics over a period of time as a baseline to pinpoint abnormal dynamics. I-seismograph reports ASes that were affected the most as well as AS paths segments that surged significantly during an incident. 

Testart \emph{et al.} \cite{Testart:2019:Serial:Hijackers} characterized the distinctive features of ASes that have been constantly reported for hijacking IP blocks for malicious purposes, i.e, serial hijackers. They relied on this knowledge to train a classifier to identify ASes with similar characteristics to those of serial hijackers. Among the most important features used for the classification task, they mentioned the intermittent AS presence and volatile prefix origination behavior. 

There are also other studies relying on signal processing and machine learning to detect BGP anomalies. Ganiz \emph{et al.} \cite{Ganiz:2006:BGP:Anomalies:Higher-Order:PathAnalysis} proposed a supervised learning method to distinguish between different anomalies in BGP traffic analyzing patterns in higher order paths. Mai \emph{et al.} \cite{Mai:2008:BGP:Anomalies:Wavelet} presented a framework to achieve both temporal and spatial localization of BGP anomalies based on wavelet analysis and clustering. Prakash \emph{et al.} \cite{Prakash:2009:BGP-lens} found patterns and anomalies in BGP updates, including self-similarity and power-law like tail distribution of updates. They used these patterns to find anomalies using median filtering. Deshpande \emph{et al.} \cite{Deshpande:2009:Online:BGP:Instability} introduced a mechanism for capturing changes in features extracted from BGP update messages using statistical pattern recognition. They found that features such as AS path length and AS path edit distance were useful in characterizing the behavior of the Internet under stress. 

More recent work has focused on the use of deep learning and time series feature extraction for BGP anomaly detection. Cheng \emph{et al.}\cite{Cheng:2016:Multi:Scale:LSTM:BGP:Anomaly:Detection} proposed a supervised multi-scale Long Short-Term Memory (LSTM) model for detecting BGP anomalies. They used worm attacks time series to train and test their method based on modeling multi-dimensional time sequences. Cheng \emph{et al.} \cite{Cheng:2018:LSTM:BGP:Anomaly:Clasification} introduced a multi-scale LSTM model that obtains temporal information at multiple scales using the Discrete Wavelet transform. They tested their method on previous worm attacks and a single path leak. McGlynn \emph{et al.} \cite{Mcglynn:2019:BGP:Anomalies:Deep:Learning} used an auto-encoder for encoding a high dimensional representation of benign routing data. Significant deviations between inputs and the output of the auto-encoder are reported as anomalies. They used their approach for detecting Multiple-Origin AS conflicts and prefix hijacks. Fonseca \emph{et al.} \cite{Fonseca:2019:BGP:Feature:Extraction} developed a dataset generation tool for extracting relevant BGP update message features as well as assisting with its labeling. They focused on volume and AS-PATH features which are commonly used by BGP anomaly detection techniques. They used their tool for generating datasets of features from different past incidents, the latest dated in 2016.

Compared with all the studies mentioned above, the present paper is unique in that it leverages the fact that certain BGP incidents, including large-scale and those that lead to traffic interception, show characteristics of highly significant burstiness. We borrow ideas from human dynamics to compute burstiness. We leverage this observation to propose a statistic-based anomaly detection method (based on the intensity of inter-arrival times) to show that it can detect both a variety of incidents with high recall and better precision than the volume-based baseline. In contrast with the work in \cite{Zhang:2004:Anomalous:Routing:Dynamics}, we did not make assumptions that require the use of training data to run the analysis. In addition, we extend the analysis for more than a limited number of prefixes and collectors. We use manually verified ground truth metadata of incidents and apply a methodology to compute the performance of the detection task. This new method found, for the first time, a quantifiable way to distinguish between normal and strange update dynamics based on burstiness. As opposed to the works in \cite{Chen:2016:Large:Scale:BGP:Detection, Zhang:2017:Seismograph}, we only use update timestamps and do not require either aggregated information from collectors nor metadata from updates. In addition, in contrast with the work in \cite{Testart:2019:Serial:Hijackers}, our focus is on the both the detection of anomalous routing events and the perpetrator instead of only malicious ASes. 

Anomaly detection methods are widely used in the networking community. We also detail recent related work in anomaly detection in applications beyond BGP that use a related approach. Lakhina \emph{et al.}~\cite{Lakhina:2004:Diagnosing:Network:Anomalies} proposed a method to detect anomalies in network traffic. Their method uses principal component analysis to distinguish between normal (predictable) and abnormal (noisy) components. They evaluated their method in network-wide traffic. Li \emph{et al.}~\cite{Li:2006:Detection:Anomalies:Sketch:Subspaces} developed a method to enable the identification of the underlying cause of network traffic anomalies. Their method is based on traffic sketches (random aggregation of IP flows) to identifyIP flows(s) that are the cause of the anomalies. Liu \emph{et al.}~\cite{Liu:2015:Opprentice} introduced Opprentice. Opprentice is a detection framework that applies supervised machine learning to automatically combine and tune diverse anomaly detection methods with the aim of optimizing operators' accuracy preference. Zhou \emph{et al.}~\cite{Zhuo:2017:Understanding:Packet:Corruption:Data:Centers} developed CorrOpt. CorrOpt is a system to mitigate packet corruption in data center networks based on an optimization algorithm. CorrOpt lowers corruption losses by three to six orders of magnitude by intelligently selecting corruption links to disable while satisfying capacity constraints. Hu \emph{et al.}~\cite{Hu:2020:Cablemon:Improving:Reliability:Cabl:Broadband:Networks} designed CableMon. CableMon is a system that applies machine learning to proactive network maintenance data to improve the reliability of cable broadband networks by ticketing predictions. CableMon generates statistical features from time series data and customer trouble tickets to infer abnormal thresholds for these generated features. CableMon prediction accuracy is four times higher than the existing public-domain tools.




\subsubsection{Other Prior Works Related to the Present Study}

Lad \emph{et al.} \cite{Lad:2006:PHAS} proposed a Prefix Hijack Alert System (PHAS). PHAS relies on the idea of finding unique prefixes simultaneously originating from multiple ASes---also referred to as Multiple Origin AS conflicts. Once these conflicts are detected, this method filters false positives using additional information from the network operators, e.g., checking announcements of similar prefixes from different ASes that belong to the same organization. Hu and Mao \cite{Hu:2007:Real:Time:Prefix:Hijacking} proposed a framework that launches data-plane probes only when anomalous update messages are received. This system was intended to be used as customized software installed in the routers. Khare \emph{et al.} \cite{Khare:2012MOAS} focused on correlating suspicious route announcements with past network announcements. This method can detect anomalies that have a huge impact, i.e., announcements that pollute a considerable number of paths. Shi \emph{et al.} \cite{Shi:2012:Argus} introduced Argus. Argus is an automated system that detects prefix hijacking and deduces the origin of the anomaly. Argus is based on pervasively correlating control- and data-plane data during a given time period to detect anomalies including sub-prefix hijacks. Schlamp \emph{et al.} \cite{Schlamp:2016:HEAP} introduced HEAP. HEAP relies on the idea of processing update messages to find malicious hijacked prefixes and then scanning the network to find SSL/TLS-enabled hosts. These enabled hosts allow the comparison of public keys prior and during an event, which is the basis for detection of subprefix hijacks. Sermpezis \emph{et al.} \cite{Sermpezis:2018:Artemis} proposed ARTEMIS. ARTEMIS is an AS self-operated detection system that exploits local configuration and real-time BGP data from public monitoring services. ARTEMIS provides protection from different types of attacks, including man-in-the-middle traffic manipulation, within a minute of detection. 
For comprehensive surveys on BGP anomaly detection and mitigation methods, we refer the reader to prior works \cite{butler2010survey, Al-Musawi:2017:BGP:Anomaly:Detection:Techniques:Survey, Mitseva:2018:State:BGP:Affairs}.

%

\section{Methods} \label{Methods}

To provide indicators of BGP anomalies, we leverage the statistic-based anomaly detection method SRI NIDES used in the intrusion detection context \cite{Javitz:1993:NIDES}. Essentially, our method considers route announcements as signals with expected patterns of behavior and detects deviations from the expected patterns. Our focus is on inter-arrival times rather than the specific content of the announcements themselves. We show that claiming illegitimate ownership of a fraction of the Internet requires transmitting correspondingly bursty announcements causing perturbations in the patterns of route announcements.


\subsection{Threat model} \label{Threat model}

We assume that route updates from the attacker will be indistinguishable from others emerging from the AS, thus we assume that none of the announcements can be trusted and the attacker can send out announcements at will. Note that in this threat model, the hijacker has no control over BGP update messages and traffic that originate outside of its domain.  

Our focus is on hijacks that use invalid announcements for an exact target prefix $p$. These incidents may or may not lead to traffic interception. Concretely, let AS1 be the legitimate owner of prefix $p$ and AS2 be the hijacker AS. The announcement $\{$AS$1- p\}$ is a BGP announcement for prefix $p$ with AS-PATH $\{ $AS$1\}$, which is the legitimate owner of the prefix. In contrast, during a hijack the hijacker AS2 announces an invalid origin as its own, to a prefix that it is not authorized to originate, i.e., $\{$AS$2 - p\}$. We focus on announcements leading to hijacks that have an invalid origin. We did not considerer other scenarios in which the hijacker can announce a more specific prefix or forge the AS-PATH \cite{Cho:2019:BGP:Hijacking:Classification}. Note that, however, this simple configuration of attack has been proven to be successful for conducting prefix hijacking and subsequent traffic interception. This can be done by forwarding the hijacked traffic along its existing valid route to the legitimate owner. As it was estimated in \cite{Ballani:2007:Prefix:Hijacking:Interception}, Tier 3 ASes and beyond can hijack traffic up to $31\%$ of ASes and intercept traffic up to $17\%$ of ASes.

\subsection{Data sources} \label{Data sources}

\subsubsection{Ground truth} \label{Ground truth}

We consider two types of routing incidents. The first type corresponds to large-scale incidents because of their impact to the Internet and the sheer number of compromised prefixes. The second type corresponds to BGP Man-In-The-Middle (MITM) incidents in which the AS attacker hijacks traffic but ultimately sends it to the victim \cite{Hepner:2009:BGP:MITM:BlackHat}.  

These exceptional selected routing incidents have not previously received detailed academic analysis, so we could not know their patterns of diffusion in advance. We obtained the details of each of the events from BGPMon \cite{BGPMon:Blog}, Oracle Dyn \cite{Dyn:2019:Security:Blog}, and Ars Technica \cite{Ars-Technica:IT:Blog}. Note that these incidents have been analyzed and corroborated from different sources. Details about the incidents and their respective dates and times are listed below. Events are listed in chronological order within each type. We used the following incidents as large-scale BGP hijacks case studies.


\noindent \textbf{An Indonesian ISP hijacks the world.} On April 2, 2014, starting at 18:26 UTC, AS4761 (Indosat), one of the largest telecommunications providers in Indonesia, announced more than $320,000$ IP prefixes belonging to other networks. Indosat announced roughly two-thirds of the entire Internet address space \cite{Toonk:Indonesia-Hijack, Zmijewski:IndonesiaHijack}. A large fraction of the hijacked prefixes belonged to Akamai, which is one of the largest Content Delivery Networks. This incident lasted approximately for $2.9$ hours until 21:15 UTC. Traffic continued to be delivered; however, the path of the traffic was significantly altered. 

\noindent \textbf{Global collateral damage of the Telecom Malaysia leak.} On June 12, 2015, starting at 08:43 UTC, AS4788 (Telecom Malaysia) announced about $179,000$ IP prefixes to Level 3 (the largest transit AS) \cite{Toonk:TelecomMalasya-Leak, Madory:2015:Dyn:Malaysia:Leak}. Level 3 accepted these announcements and then propagated the routes to their peers and customers around the world. Because Telecom Malaysia is a customer of Level 3, the routes announced by Telecom Malaysia were identified as a preferred delivery route for Level 3. This event caused significant packet loss and Internet service degradation around the world. Level 3 suffered a significant blackout from the Asia pacific region and the rest of the world. Note that this was a leak, so the data were not delivered after being transmitted to Telecom Malaysia. This incident lasted approximately $2.7$ hours. At around 10:40 UTC, there were slowly observed improvements, and by 11:15 UTC the errors in the Routing Information Base (RIB) \cite{rfc4271} began to be resolved. 

\noindent \textbf{Large-scale BGP hijack in India.} On November 6, 2015, starting at 05:52 UTC, AS9498 (Bharti Airtel Ltd.) claimed the ownership of about $16,123$ IP prefixes. These addresses corresponded to more than $2,000$ unique ASes \cite{Toonk:India-Hijack, Murphy:2015:India:Hijack:NANOG}. This event became widespread because two large ASes (Cogent Communications and GlobeNet Cabos Submarinos S.A.) accepted and propagated these routes to their peers and customers. Legitimate owners of the prefixes included Akamai, Tata Communications, and Apple Inc. This incident lasted approximately $8.9$ hours until 14:40 UTC.

\noindent \textbf{Small Nigerian ISP leak takes Google down.} On November 12, 2018, starting at 21:13 UTC, AS37282 (MainOne) leaked 212 prefixes belonging to Google \cite{Naik:2018:Google:Leak:ThousandEyes, Goodin:2018:Google:Leak:ArsTechnica}. This leak caused Google's traffic to drop at AS4809 (China Telecom), who improperly accepted the routes and announced them world-wide. This incident affected services such as Google Workspace (formerly G suite), Google Search, and Google Analytics. The redirection came in five distinct waves over a 74 minute period lasting until approximately 22:27 UTC.

We used the following incidents as BGP MITM case studies.

\noindent \textbf{Belarusian traffic redirection.} On February 27, 2013, starting at 08:01 UTC,\footnote{A contact from Oracle Dyn confirmed these details.} AS28849 (GlobalOneBel) redirected global traffic in a sequence of events lasting from a few minutes to several hours in duration \cite{Cowie:2013belarus}. Redirections were claimed to happen almost on a daily basis through February. The set of victims were reported to be changing constantly including financial institutions, governments, and network providers. In our analysis, we focus on an incident in which traffic that was intended to go from Guadalajara, Mexico to Washington, DC went through Belarus.

\noindent \textbf{Icelandic traffic redirection.} On July 31, 2013, starting at 07:36 UTC, AS48685 (Opin Kerfi) began announcing the ownership of $597$ prefixes of a large VoIP provider (one of the largest facilities-based providers of managed services in the U.S.) \cite{Cowie:2013belarus}. This was one of the $17$ incidents during the period July 31 to August 30. These hijacks affected different victims in other countries sharing a common pattern. Particularly, false routes were sent to the hijacker's peers in London leaving clean paths to destinations in North America. This was with the aim of redirecting traffic back to its original destination. In our analysis, we focus on an incident in which traffic between two destinations in Denver, Colorado went through Iceland.

\noindent \textbf{Russian telecom hijacks financial services.} On April 26, 2017, starting at 22:36 UTC, AS12389 (Rostelecom) started to announce $50$ prefixes from $37$ different ASes \cite{Goodin:2017:Russian:BGP:MITM:Attack:ARS:Technica, Toonk:2017:Russian:BGP:MITM:Cisco}. This incident affected prefixes from several financial institutions including Visa, MasterCard, and more than two dozen other institutions, including security companies, such as Symantec. During this incident, the traffic of these institutions was briefly routed through the Russian telecom before being sent to its original destination. This incident lasted approximately seven minutes until 22:43 UTC.

\subsubsection{BGP data} \label{BGP data}

After obtaining each of the incident details, we collected historical BGP updates (announcements and withdrawals) using BGPStream.\footnote{Available at https://bgpstream.caida.org/} Update timestamp accuracy is one second. BGPStream provides an open-source software framework for the analysis of historical and real-time BGP data \cite{Orsini:2016:BGPStream}. BGPStream extracts data directly from route collectors. A route collector (collector, hereafter) is a host running a collector process. The collector emulates a router that establishes BGP peering sessions with BGP routers, known as feeders. 

There are two popular projects running route collector processes, RouteViews \cite{Meyer:RouteViews} and RIPE RIS \cite{NCC:RIPENCC}. At the time of this writing, RouteViews and RIPE RIS operate $29$ and $24$ collectors respectively which peer with hundreds of feeders \cite{Gregori:2012:BGP:Route:Collector}. We acknowledge that there are other sources of BGP data, including network operators, other route collector projects such as BGPmon \cite{Yan:2009:BGPMon} (from Colorado State University)\footnote{This refers to the free BGP monitoring service available at \href{https://www.bgpmon.io/}{https://www.bgpmon.io/}} and Packet Clearing House (PCH) \cite{PCH:Collector:Project}. BGPmon is a distributed system that monitors BGP data by peering with multiple ASes. It provides access to real-time data that is available in XML format. PCH operates route collectors at different Internet Exchange Points around the world. PCH made this data publicly available on its website. BGPmon and PCH provides access to a limited number of feeders when compared to RouteViews and RIPE RIS \cite{Gregori:2015:AS-level:Incompleteness}. Among the last two, previous research has shown that there is a considerable overlap between the measurements from RouteViews and RIPE RIS projects \cite{Chen:2009aeyeshots}. However, as pointed out in \cite{Gregori:2015:AS-level:Incompleteness}, RouteViews provides the more complete view of the Internet in terms IP prefix coverage. This was estimated by showing that RouteViews collectors peer with more full feeders (i.e., those feeders that announce an IPv4 (IPv6) space closer to the the full Internet IPv4 (IPv6) space currently advertised). Therefore, we only collected BGP updates from RouteViews.


Our data collection is based on a subset of BGP updates that cover the time before, during, and after selected incidents. We collected approximately seven days of observations around the start date of each of them. The purpose of collecting data over this time period is to be able to distinguish between regular and anomalous behavior. This time period is long enough to capture the duration of each incident. This means that when we collected data during this time period, we can guarantee that we will monitor the impact of the incident during the selected time frame. 

We also verified that the incidents studied here were very unlikely to be generated by session resets. Session resets occur when a BGP session between a collector and feeder is re-established. Resets require the complete routing table of the feeder to be sent to the collector, generating a large number of announcements. We used the algorithm in \cite{Cheng:2011:Identifying:BGP:Routing:Tables} to verify that these incidents were not originated by session resets.

\subsection{Burstiness of announcements} \label{Burstiness of announcements}

Burstiness refers to the tendency of certain events to occur in groups of relatively high frequency, i.e., short inter-arrival time intervals, followed by periods of relatively infrequent events \cite{Harang:2017:Burstiness:Intrusion:Detection, Xu:2018:Modeling:Cyber:Hacking:Breaches}. Let $X_{\alpha \to \beta} = \{ X_{\alpha \to \beta}(t)\}, t = 0, 1, \ldots, n-1$ be a time series of time-stamped announcements originated by AS$\alpha$ and received by collector $\beta$. Let $\tau_{\alpha \to \beta}$ be a random variable that represents the time interval between consecutive announcements so that $\tau_{\alpha \to \beta}$ takes values in $\{ X_{\alpha \to \beta}(1) - X_{\alpha \to \beta}(0), X_{\alpha \to \beta}(2) - X_{\alpha \to \beta}(1), \ldots, X_{\alpha \to \beta}(n-1) - X_{\alpha \to \beta}(n-2) \}$. Mathematically, burstiness can be characterized by analyzing the inter-arrival time distribution $P(\tau_{\alpha \to \beta})$. As proposed in \cite{Goh:2008:Burstiness:Complex:Systems}, the inter-arrival distribution can be characterized by a burstiness factor defined by $B=\frac{\sigma - \mu}{\sigma + \mu}$. Here, $\sigma$ and $\mu$ denote the standard deviation and mean of the inter-arrival time distribution. Note that the burstiness has a value of $-1$ for $\sigma=0$, which means regular time intervals. It has a value of $0$ for $\sigma=\mu$ in the case of random time intervals. Finally, it has a value of $1$ for $\sigma \to \infty$ and a finite $\mu$ in the case of a highly bursty time series of announcements.  

Note that the original formulation of burstiness depends on the number of events used to describe the inter-arrival time distribution, i.e., the value of $n$. We used a modified version of burstiness that deals with the finite size effects of collected announcements while maintaining the same scale defined in~Eq.~\eqref{eq:Burstiness} \cite{Kim:2016:Burstiness:Finite:Sequence}. In our analysis, we computed the burstiness of ASes that sent at least five announcements during the period of study. 
\begin{equation}
B(n) = \frac{\sqrt{n+1} - \sqrt{n-1}  + (\sqrt{n+1} + \sqrt{n-1})B}{\sqrt{n+1} + \sqrt{n-1} - 2 + (\sqrt{n+1} - \sqrt{n-1} -2)B}
\label{eq:Burstiness}
\end{equation}

\subsection{Detection method} \label{Detection method}

We leverage the measure of inter-arrival times as received by the collectors to compute a measure of intensity based on the burstiness of announcements. This measure was originally used in the context of intrusion detection in \cite{Javitz:1993:NIDES}. Let $Q_{\alpha \to \beta}$ be the number of announcements sent by AS$\alpha$ and received by collector $\beta$, exponentially weighted. This means that more current announcements have a greater impact in its computation, i.e., short inter-arrival times. The value of $Q_{\alpha \to \beta}$ is computed using the recursive formula 
\begin{equation}
Q_{\alpha \to \beta}(t) = 1 + 2^{-r \Delta}Q_{\alpha \to \beta}(t-1). 
\label{eq:Q computation}
\end{equation}
Here, $r$ is the decay factor and $\Delta = X_{\alpha \to \beta}(t) - X_{\alpha \to \beta}(t-1)$ is the inter-arrival time between consecutive announcements. The decay factor $r$ determines the half-life of $Q_{\alpha \to \beta}(t)$. Large values of $r$ imply that the value of $Q_{\alpha \to \beta}(t)$ is more influenced by more recent announcements. Smaller values of $r$ imply that the value of $Q_{\alpha \to \beta}(t)$ will be more heavily influenced by announcements in the distant past. In accordance with previous studies in \cite{Labovitz:2000:Delayed:Routing:Convergence, Zhang:2004:Anomalous:Routing:Dynamics}, we verified that most inter-arrival times of announcements are less than $300$ seconds (the $99$th percentile for most of the collectors). We then use $300$ seconds as the half-life value to capture most of routing dynamics. Then the decay factor is set to be $r = 1/300$.

\subsubsection{Time series anomaly detection} \label{Time series anomaly detection}

Detection focuses on identifying anomalous observations in the time series of $Q_{\alpha \to \beta}$. We do that by forecasting the time series and using predictions as the basis to identify abnormal behavior. This allows us to deal with trends in the time series. This procedure has been used before for anomaly detection in time series \cite{Nakano:2017:Generalized:EMA:Anomaly:Detection, Davis:2019:LSTM:Anomaly:Detection:Extreme:Value}. In particular, we used predictions and the standard deviation of the predicted time series as a proxy for the error to pinpoint anomalies. We used two different models for time series prediction. The first one is based on exponential moving average (EMA), which is a statistical-based method \cite{Hyndman:2008:EMA}. We used EMA and its standard deviation as the mean and standard deviation estimators, respectively. EMA is well suited to work with unevenly spaced (also called unequally or irregularly spaced) time series data \cite{Muller:1991:EMA} as it is the case in this analysis. In addition, it has been shown that traditional statistical methods, such as EMA, are more accurate and less computationally expensive than machine learning-based methods, including those relying on neural networks, at least for forecasting purposes \cite{Makridakis:2018:Stats:ML:Forecasting}. The second one is based on a Long Short-Term Memory (LSTM) network, which is a deep learning-based method \cite{Hochreiter:1997:LSTM}. We used LSTM predictions and its standard deviation as the mean and standard deviation estimators, respectively. We described EMA and LSTM in detail below.

\noindent \textbf{EMA}. EMA computes estimations based on weighted averages of past observations. EMA weights follow an exponential decay. That means that more recent observations have more weight than past observations. The recursive equations (for efficient computation in a data stream) for the mean, variance, and standard deviation of EMA are based on \cite{Finch:2009:EMA} and defined by
\begin{equation} \label{eq:EMA}
\mu(t) = a y(t) + (1-a)\mu(t-1)
\end{equation}
where $\mu(t)$ and $y(t)$ are the mean estimate and the value of the time series at time $t$, respectively. The parameter $a \in [0,1]$ is the weighting decrease. A higher value of $a$ discounts older observations faster. EMA can be also parametrized using the window length $\omega$. The relationship between $a$ and $\omega$ is given by $a = \frac{2}{1 + \omega}$. 
Similarly, the variance $S(t)$ and standard deviation $\sigma(t)$ estimates at time $t$ are defined by
\begin{eqnarray} \label{eq:EMS}
\sigma^2(t) &=& S(t) = (1 - a)(S(t-1) + a(y(t)-\mu(t-1))^2) \nonumber \\
\sigma(t) &=& \sqrt{S(t)} 
\end{eqnarray}
We used $\omega = 200$ as the estimator for the window length because it is the lowest value where the root mean square error between the empirical observations and the EMA begins to flatten. This is consistent across collectors.

\begin{figure}[!htbp]
\setlength\abovecaptionskip{-0.5\baselineskip}
\centering
\includegraphics[width=0.75\columnwidth]{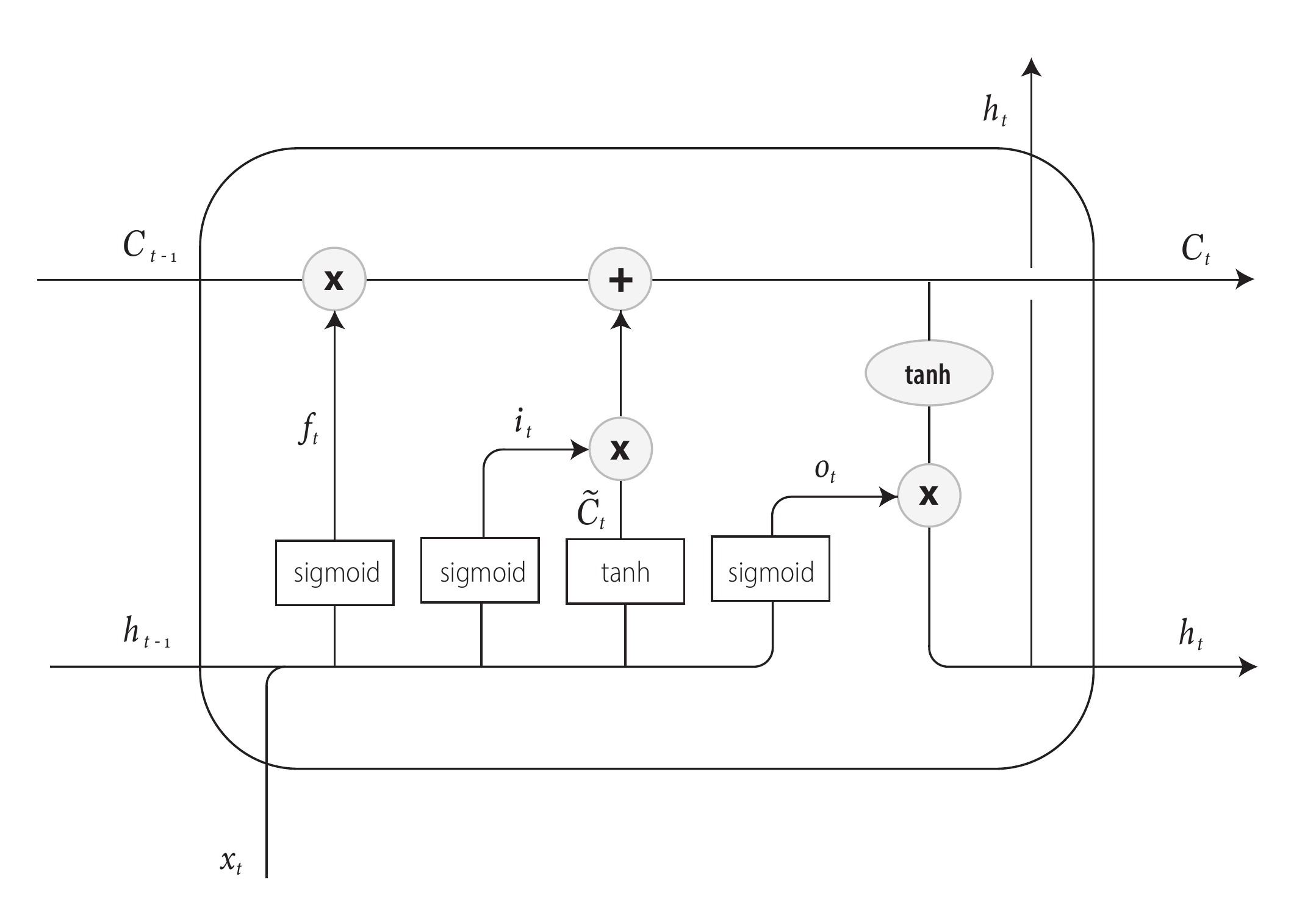} 
\caption{Structure of an LSTM cell.}
\label{fig-lstm-cell}
\end{figure}

\noindent \noindent \textbf{LSTM}. An LSTM network is a recurrent neural network architecture specifically designed to address the vanishing gradient problem, i.e., a backpropagation error that either growths or decays exponentially. This makes LSTMs ideal to model long-term dependencies. An LSTM network can be composed of multiple layers, and each layer features a set of recurrently connected blocks, known as cells. Each cell has three multiplicative units also known as gates, i.e., forget, input, and output gates. They provide the functionality to reset, write, and read the cell. Figure~\ref{fig-lstm-cell} shows the structure of an LSTM cell. 
The outputs of the gates are calculated as follows:
\begin{eqnarray*}
f_{t} &=& \text{sigmoid}(W_{f} [h_{t-1}, x_{t}] + b_{f}) \\
i_{t} &=& \text{sigmoid}(W_{i} [h_{t-1}, x_{t}] + b_{i}) \\
o_{t} &=& \text{sigmoid}(W_{o} [h_{t-1}, x_{t}] + b_{o}) \\
\end{eqnarray*}
The new cell state $C_{t}$ is updated using: 
\begin{eqnarray*}
\tilde{C}_{t} &=& \tanh(W_{c} [h_{t-1}, x_{t}] + b_{c}) \\
C_{t} &=& f_{t} \circ C_{t-1} + i_{t} \circ \tilde{C}_{t}
\end{eqnarray*}
And the output to the next cell is:
\begin{eqnarray*}
h_{t} &=& o_{t} \circ \tanh(C_{t}),
\end{eqnarray*}
where $W_{*}$ and $b_{*}$ are the weight matrices and bias vectors, respectively. We divided each time series dataset per collector into a training and a validation dataset. We used the first day of observations for training. We assumed that the observations during the first day are free from anomalies given that the incidents occurred approximately in the half of the seven-day period, i.e., the model learns the normal behavior of the time series. We selected the hyperparameters, including network architecture and batch size, using cross-validation on the validation dataset. We obtained consistent results across collectors using one LSTM layer, four cells, and a batch size of 64. The output layer is a fully connected dense layer with linear activation. We used the Adam optimizer and the mean square error as the objective function. We used Keras with the TensorFlow backend for model implementation.

\subsubsection{Detection criterion} \label{Detection criterion}

We pinpointed observations in the time series $Q_{\alpha \to \beta} (t)$ based on how far they are from the predictions. We calibrated the distance threshold from the predictions using the standard deviation. The parameter $\delta$ controls how many standard deviations are considered to report an anomaly. The results presented in this work are based on using $\delta=2$. When we set this threshold for the EMA/LSTM predictions, we can detect anomalies with an error rate of $4.5\%$ since $P(\mbox{EMA/LSTM} - 2\delta <= Q_{\alpha \to \beta} (t)  <= \mbox{EMA/LSTM} + 2\delta) \approx 95.5\%$. The values of $r$, $\omega$, and $\delta$ may be tuned for detection purposes. The complete pseudocode for the detection algorithm can be found in Algorithm~\eqref{alg-Algorithm-Event-Detection}.
\begin{algorithm}[htp!] 
\scriptsize
\caption{Event-Detection ($X_{\alpha \to \beta}$, $r$, $\omega$, $\delta$)} 
\label{alg-Algorithm-Event-Detection}
\begin{algorithmic}[1]
\State $Q_{\alpha \to \beta} \gets \{ 0 \}$ \Comment{Number of announcements}
\State $\Psi \gets \{ 0 \}$ \Comment{EMA/LSTM}
\State $\Sigma \gets \{ 0 \}$ \Comment{EMA/LSTM standard deviation}
\State $\hat{A} \gets \{ \}$ \Comment{Anomalous timestamps}
\For{$t$ in $\{1, \ldots, n-1 \}$}
\State $Q_{\alpha \to \beta} \gets Q_{\alpha \to \beta} \cup \{  1 + 2^{-r \Delta}Q_{\alpha \to \beta}(t-1) \}$ using Eq.~\eqref{eq:Q computation} 
\State $\Psi \gets \Psi \ \cup$ EMA/LSTM($Q_{\alpha \to \beta}(t), \Psi(t-1), \omega$) using Eq.~\eqref{eq:EMA} 
\State $\Sigma \gets \Sigma  \ \cup$ EMA/LSTM std($Q_{\alpha \to \beta}(t), \Psi(t-1), \Sigma(t-1), \omega$) using Eq.~\eqref{eq:EMS} 
\If{$Q_{\alpha \to \beta}(t) >= (\Psi(t)  + \delta \Sigma(t)$)}
\State $\hat{A} \gets \hat{A} \cup \{ t \}$
\EndIf 
\EndFor
\State \textbf{return} $\hat{A}$
\end{algorithmic}
\end{algorithm}

\subsection{Detection evaluation} \label{Detection evaluation}

We compute the performance of the detection method by correlating significant deviations in the smoothed time series and the ground truth defined in Section~\ref{Ground truth}. Note that the proposed method is unsupervised and that labeling is performed only to generate the ground truth for evaluating the performance of the proposed method. We compare the proposed method against the baseline of volume of announcements. To do so, we partition the time domain under study into equally binned size intervals. We adjust the detection resolution by changing the length of the intervals denoted by $m$. Let $N$ be the number of times at which detection is assessed using consecutive intervals on length $m$. Let $E \subseteq \{1, 2, \ldots, N\}$ represents the time intervals at which one event occur based on the ground truth. Let $\hat{E} \subseteq \{1, 2, \ldots, N\}$ represents the time time intervals at which at least one event is reported based on the detection method. The performance is then measured based on $E$ and $\hat{E}$. This procedure has been used before to report performance of event detection methods in \cite{Moriano:2017:Insider:Treat:Graph:Mining, Moriano:2019:Community:Event:Detection}. 
We compared the performance of the proposed method and the volume baseline by counting the number of time intervals that are true positives (TP), false positives (FP), false negatives (FN), and true negatives (TN) in the detection task. 
Precision quantifies the proportion of reported intervals that were correctly detected. That means that erroneously reporting a considerable number of intervals produces low precision. It is quantified as $\frac{TP}{TP + FP}$. Recall quantifies the proportion of actual anomalous intervals that were correctly detected. That means that having a considerable number of intervals that were not reported produces low recall. It is quantified as $\frac{TP}{TP + FN}$. F1 score quantifies the balance between precision and recall through the harmonic mean. It is quantified as $2\frac{\text{precision} \times \text{recall}}{\text{precision} + \text{recall}}$.

\section{Results} \label{Results}

In this section, we analyze the previously described BGP incidents. Due to space constraints, here we present the analysis of the following incidents: Indosat, Telecom Malaysia, and Rostelecom. For the analysis of the remaining incidents: Bharti Airtel Ltd., MainOne, GlobalOneBel, and Opin Kerfi, please refer to the Supplement, Section~\ref{Remaining incidents appendix}. We analyzed the views from several data collectors at various locations around the world. Table~\ref{Table:Collector_placement} shows the geographical location and the date of the first dump of the collectors used in this study.\footnote{Collectors' location and date of the first dump were obtained from \href{http://www.routeviews.org/routeviews/index.php/collectors/}{RouteViews} and \href{https://bgpstream.caida.org/data}{BGPStream} respectively.}
\begin{table}[htp!] 
\caption{Geographical location and date of first dump of collectors. Collectors are ordered in alphabetical order.} 
\centering 
\small
\tabcolsep=0.07cm
\begin{tabular}{| l | l | l |} 
\hline
\multicolumn{1}{|c|}{\textbf{Collector name}} & \multicolumn{1}{c|}{\textbf{Location}} & \multicolumn{1}{c|}{\textbf{First dump}} \\
\hline
\hline
route-views.amsix & Amsterdam, NL & 2018-07-11 23:19 \\ 
route-views.chicago & Chicago, IL, US & 2016-06-28 12:00 \\ 
route-views.chile & Santiago, CL & 2018-01-31 20:00 \\ 
route-views.eqix & Ashburn, VA, US & 2004-05-17 13:59 \\ 
route-views.flix & Miami, FL, US & 2018-01-19 16:00 \\ 
route-views.isc & Palo Alto, CA, US & 2003-11-27 02:00 \\ 
route-views.jinx & Johannesburg, ZA & 2012-07-10 00:00 \\ 
route-views.kixp & Nairobi, KE & 2005-10-07 15:44 \\ 
route-views.linx & London, GB & 2004-03-16 13:45 \\ 
route-views.napafrica & Johannesburg, ZA & 2018-02-01 02:00 \\ 
route-views.nwax & Portland, OR, US & 2014-03-20 20:52 \\ 
route-views.perth & Perth, AU & 2012-11-15 21:48 \\ 
route-views.saopaulo & Sao Paulo, BR & 2011-03-17 16:19 \\ 
route-views.sfmix & San Francisco, CA, US & 2015-04-14 20:00 \\ 
route-views.sg & Singapore, SG & 2014-06-04 15:44 \\ 
route-views.soxrs & Belgrade, RS & 2014-01-01 00:00 \\ 
route-views.sydney & Sydney, AU & 2010-08-14 02:00 \\ 
route-views.telxatl & Atlanta, GA, US & 2012-02-02 22:46 \\ 
route-views.wide & Tokyo, JP & 2003-07-01 21:29 \\ 
route-views2 & Eugene, OR, US & 2001-10-26 16:48 \\ 
route-views3 & Eugene, OR, US & 2013-11-25 10:00 \\ 
route-views4 & Eugene, OR, US & 2008-11-28 09:53 \\ 
route-views6 & Eugene, OR, US & 2003-05-03 12:29 \\ 
\hline
\end{tabular} 
\label{Table:Collector_placement} 
\end{table}
We analyzed BGP announcements and withdrawals, but the withdrawals did not affect our results, perhaps in part because the volume of withdrawals is significantly less \cite{Wang:2002:GBP:Under:Stress, Lad:2003:Analysis:BGP:Worm, Deshpande:2004:Early:Detection:BGP:Instabilities}. Our results presented here include only announcements. For the following analysis, we focused on the view of the top four collectors based on the number of feeders. For the remaining collectors, please refer to the Supplement, Section~\ref{Remaining collectors appendix}. We conduct three different but complementary analyses. 


First, we show how each of the incidents is observed from the point of view of the different collectors (Section~\ref{Collectors' disruption perception}). To do so, we measure the number of announcements (i.e., unique originated prefixes) received by the collectors and that were sent by the perpetrator AS, before, during, and after the incident. We show how using the volume of these announcements as the basis for anomaly detection may be useful to pinpoint these anomalies. This strategy, however, produces more false positives.

Second, we analyze the inter-arrival times of announcements at the collectors (Section~\ref{Inter-arrival time analysis}). We characterize these with a measure of burstiness used previously for studying human dynamics in \cite{Barabasi:2005:Origin:Bursts, Goh:2008:Burstiness:Complex:Systems}. This allows us to quantify the burstiness of announcements before, during, and after the incidents. We show that ASes involved in the reported incidents exhibit a statistically significant change in the inter-arrival pattern of their BGP announcements at the collectors. We show that for the detection of BGP incidents, the volume of messages is not enough for incident detection. In contrast, the burstiness of the announcements sent by ASes and seen for a specific collector is a better discriminator than the volume of announcements in identifying anomalous behavior. More importantly, we show that changes in burstiness correlate with occurrence of the incidents. 

Third, we detail a method for detecting anomalous announcements based on quantifying inter-arrival times of announcements received by the collectors (Section~\ref{Anomaly detection}). This is done by leveraging the observation that there is a significant change of burstiness during the incidents. This allows us to characterize the distinguishing feature that occurs during the incidents. Based on this distinguishing feature, we introduce a detection algorithm and evaluate its effectiveness using real-stream data obtained from collectors during the incidents. We compare the performance of the anomaly detection method based on bustiness and the baseline of volume. We show that for the task of detecting anomalous behavior, the method based on bustiness is able to reduce false positives considerably while correctly detecting the incidents. 

\subsection{Collectors' disruption view} \label{Collectors' disruption perception}

The reported anomalous behavior (ground truth) in each of the incidents is highlighted in the figures by the vertical dashed lines. We ranked the collectors in decreasing order by the number of feeders. In this analysis, vertical circles represent the number of unique originated prefixes. The horizontal solid line represents the value of the EMA estimate. Each horizontal gray band represents one standard deviation from the EMA using the same window length (more intense bands indicate observations that are further away from the mean, based on Algorithm~\ref{alg-Algorithm-Event-Detection}). Observations that are more than two standard deviations away from the EMA are marked with stars. 

\begin{figure}[!htbp]
\setlength\abovecaptionskip{-1.5\baselineskip}
\centering
\includegraphics[width=1.0\columnwidth]{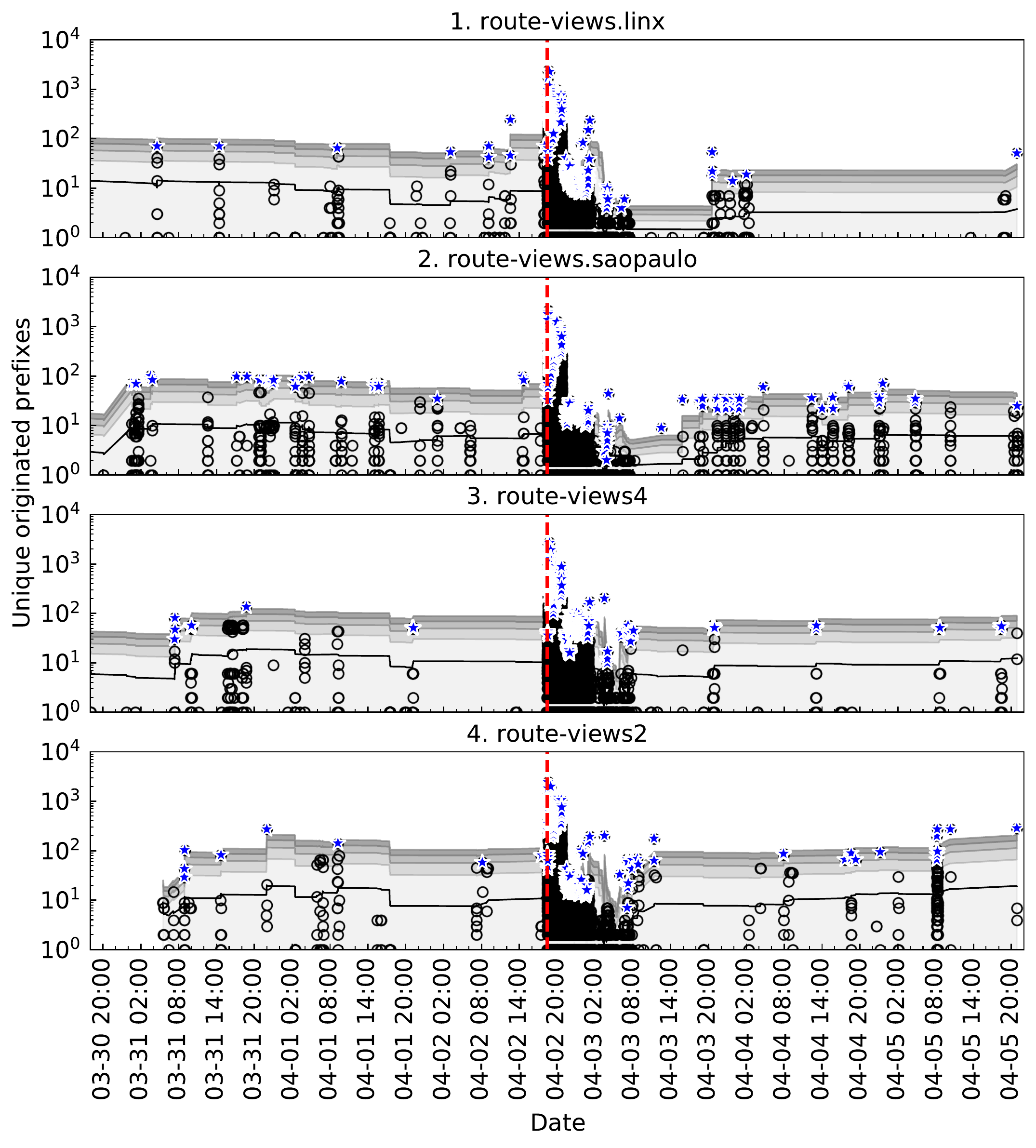} 
\caption{Time series of the number of announcements from AS4761 that collectors received before, during, and after the Indosat incident in 2014 for the top four collectors. Major ticks correspond to six-hour intervals while minor ticks correspond to two-hour intervals. }
\label{fig-time-series-announcements-indosat}
\end{figure}

\noindent \textbf{Indosat.} Figure~\ref{fig-time-series-announcements-indosat} shows the number of announcements received from the AS responsible for the incident.\footnote{Each individual circle corresponds to the raw number of unique originated prefixes at a particular timestamp. We did not bind them.} Note that two things happen. First, there is a significant increase in the number of received announcements. This increase is almost four orders of magnitude. Second, the frequency at which the announcements are received is higher than other announcements that are not close to the start of the incident. This last observation implies shorter inter-arrival times in the proximity of the incident. For these collectors, this behavior is correlated with the reported ground truth.  

\begin{figure}[!htbp]
\setlength\abovecaptionskip{-1.5\baselineskip}
\centering
\includegraphics[width=1.0\columnwidth]{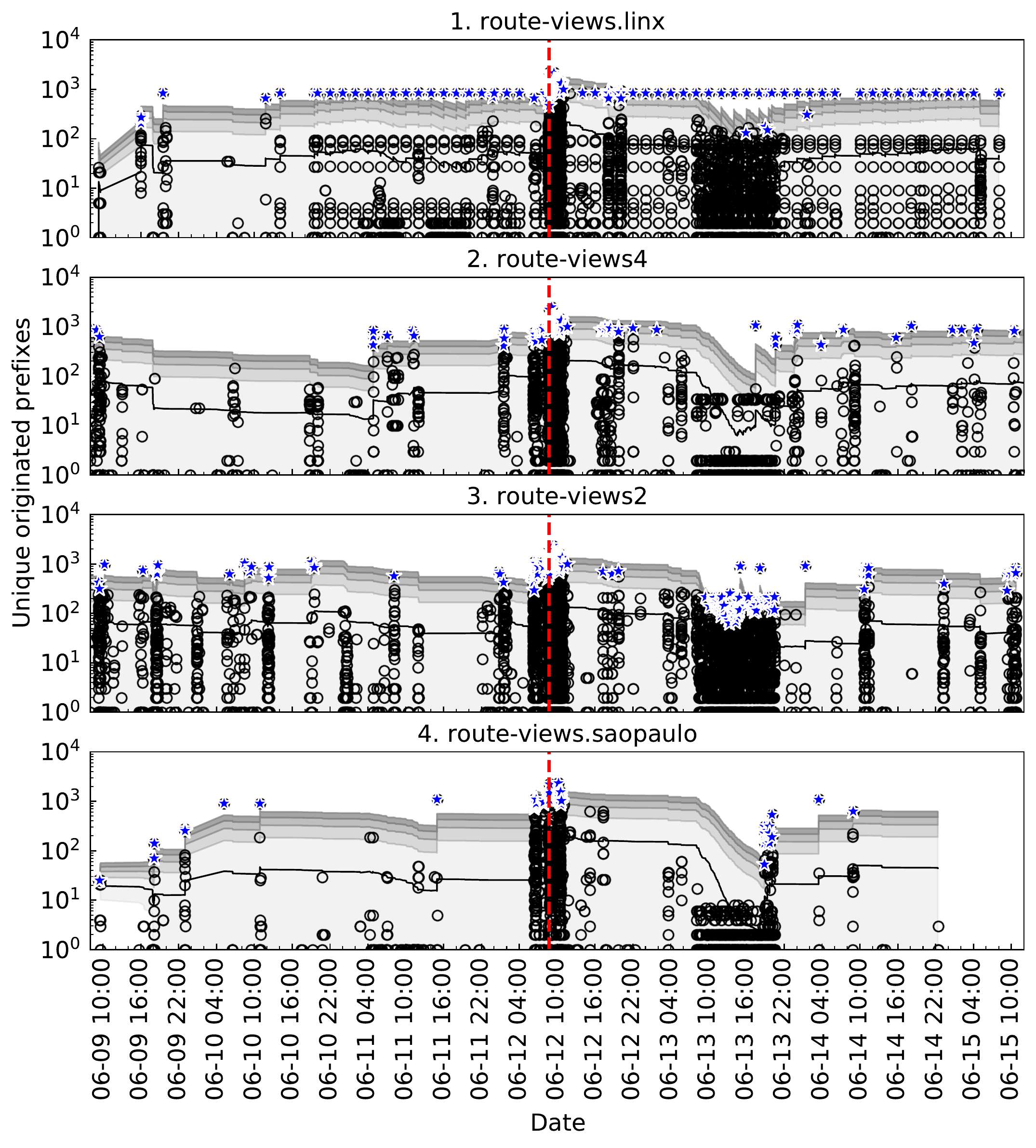} 
\caption{Time series of the number of announcements from AS4788 that collectors received before, during, and after the Telecom Malaysia incident in 2015.}
\label{fig-time-series-announcements-malaysia}
\end{figure}

\noindent \textbf{Telecom Malaysia.} Figure~\ref{fig-time-series-announcements-malaysia} shows the number of announcements received by every collector. The number of announcements increases up to four orders of magnitude. Collectors observe an increase of burstiness in announcements that is correlated with the ground truth. Even more importantly, these announcements occur highly intermittently and frequently. We also observe that there is a bursty high volume of announcements on June 13, 2015 at 08:05 UTC. We acknowledge that these announcements may be either valid because of the reestablishment of valid routes or invalid and reflecting a new incipient incident. We found that the majority of these announcements were valid. We revisit the case of invalid announcements in the Supplement, Section~\ref{Performance results with augmented ground truth appendix}.

\begin{figure}[!htbp]
\setlength\abovecaptionskip{-1.5\baselineskip}
\centering
\includegraphics[width=1.0\columnwidth]{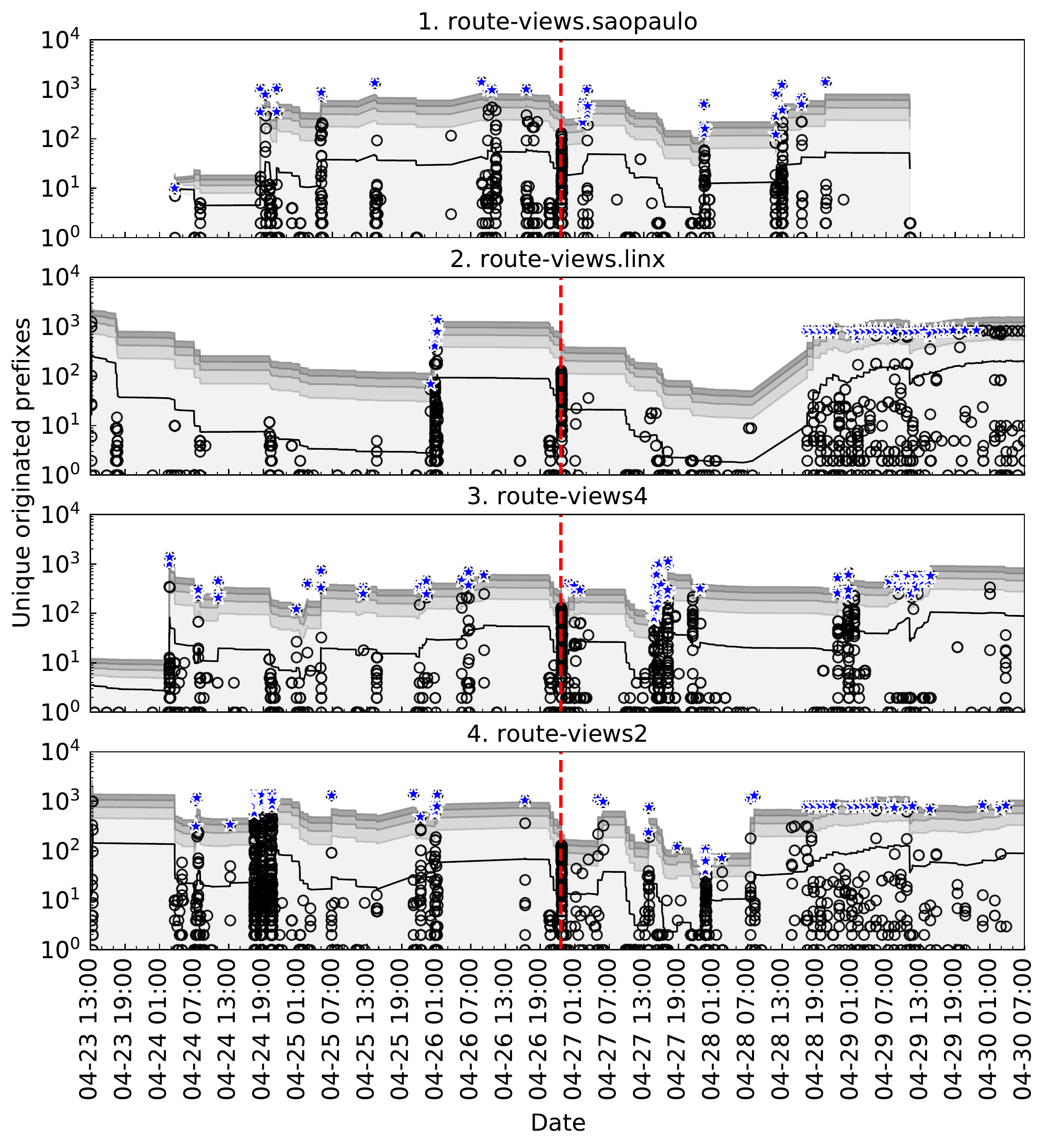} 
\caption{Time series of the number of announcements from AS12389 that collectors received before, during, and after the Rostelecom incident in 2017.}
\label{fig-time-series-announcements-russia}
\end{figure}

\noindent \textbf{Rostelecom.} Figure~\ref{fig-time-series-announcements-russia} shows an increase in the number of announcements of up to four orders of magnitude. We observed that the reported anomalies are not necessarily localized near the reported ground truth but instead across the observation period.

\subsection{Inter-arrival time analysis} \label{Inter-arrival time analysis}

There is both a significant increase in the number of arrivals of announcements during the incident and a dramatic increase in the frequency at which these announcements are received by the collectors (see Section~\ref{Collectors' disruption perception}). The following analysis reveals that the inter-arrival time of announcements as seen by the collectors exhibits a significant degree of burstiness. To ground the results, we first analyze the joint distribution of activity of each AS based on the burstiness (horizontal axis) and the number of announcements (vertical axis) during one full day of measurements around the incident. Collectors are ranked in decreasing order by number of feeders. We marked with a ``star'' the ASN that was responsible for the incident. We marked with ``squares'' the ASNs of the top three burstiests ASes. They provide a baseline for comparison. The vertical and horizontal dashed lines represent the 95$\%$ percentile of the distributions on each axis. The dark cells indicate a high concentration of ASes with a characteristic burstiness and number of announcements, which is quantified in the legend.

Second, we test if the apparent effect is real or is due to chance. In particular, we apply a Monte Carlo test in which the null hypothesis is that ASes send announcements in a bursty manner even during times where there is no evidence of BGP incidents. For this analysis, we collected time series of announcements over a full day of observations where no BGP incidents have been reported by our ground truth sources. Our ground truth sources use experts to manually check the incidents and determine potentially malicious hijacks (a similar procedure as it has been used in \cite{Cho:2019:BGP:Hijacking:Classification}). One hundred of these random time series were compiled for each collector for the top five burstiest ASes (used as a baseline) and the perpetrator AS. In each of these 100 time series, we compute the ASes associated burstiness. Here we provide the results for the top four collectors based on the number of feeders. Again, details for the other collectors are available in the Supplement, Section~\ref{Remaining collectors appendix}. 

\begin{figure}[!htbp]
\setlength\abovecaptionskip{-1.5\baselineskip}
\centering
\includegraphics[width=1.0\columnwidth]{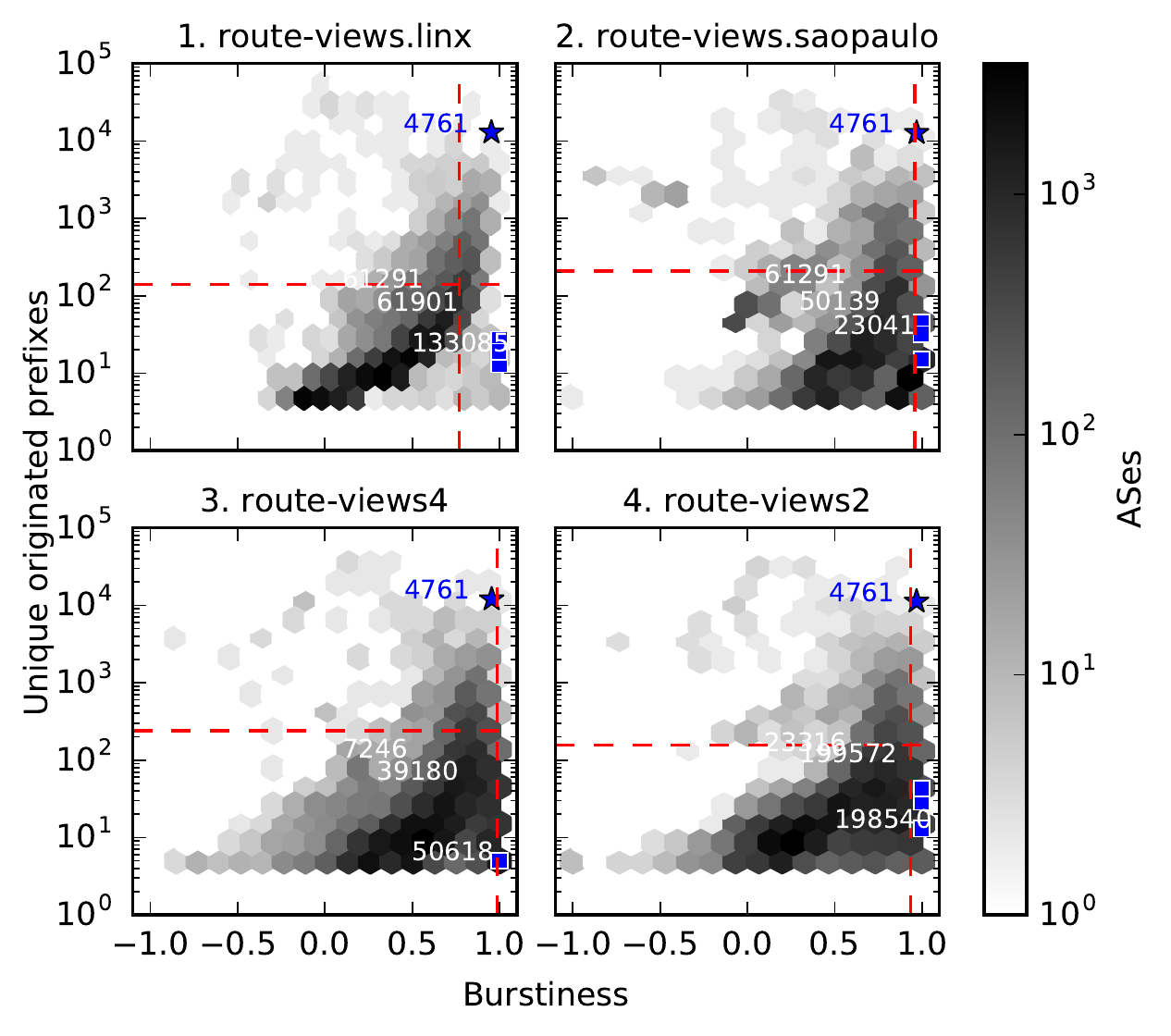} 
\caption{Joint distribution based on the burstiness (horizontal axis) and number of announcements (vertical axis) during one day interval around the Indosat incident.}
\label{fig-joint-distribution-indosat}
\end{figure}

\noindent \textbf{Indosat.} Figure~\ref{fig-joint-distribution-indosat} shows the distribution of activity of each AS. In particular, we compute their bustiness and measure the number of announcements. The AS represented by the star (i.e., Indosat) has among the highest burstiness, belonging to the first quadrant, which is even comparable to the top three burstiests ASes, marked with squares, that are placed in the fourth quadrant. Note also that there are ASes in the second quadrant that have a considerable number of announcements but lower burstiness, including, AS36947, AS41691, AS17557, AS13118, AS53062. These ASes appear consistently among the different collectors but were not reported to be involved in the incident. Conversely, those ASes in the fourth quadrant show high burstiness but not a significant number of announcements (i.e., AS61291, AS50139). Those ASes are not neighbors of Indosat (corroborated through CAIDA's AS Rank \cite{caida:2018:asrank}) nor involved in the incident. This empirical finding reveals that although the volume of announcements increases for different ASes during the incident, the actual AS involved in the incident has a distinct burstiness pattern that is correlated when the incident is reported. This observation is complementary to the works in \cite{Lad:2003:Analysis:BGP:Worm, Deshpande:2004:Early:Detection:BGP:Instabilities} in which a significant increase in the volume of announcements is used as a detection signature, as well as illustrating the benefit of including a measure of burstiness.

\begin{figure}[!htbp]
\setlength\abovecaptionskip{-1.5\baselineskip}
\centering
\includegraphics[width=1.0\columnwidth]{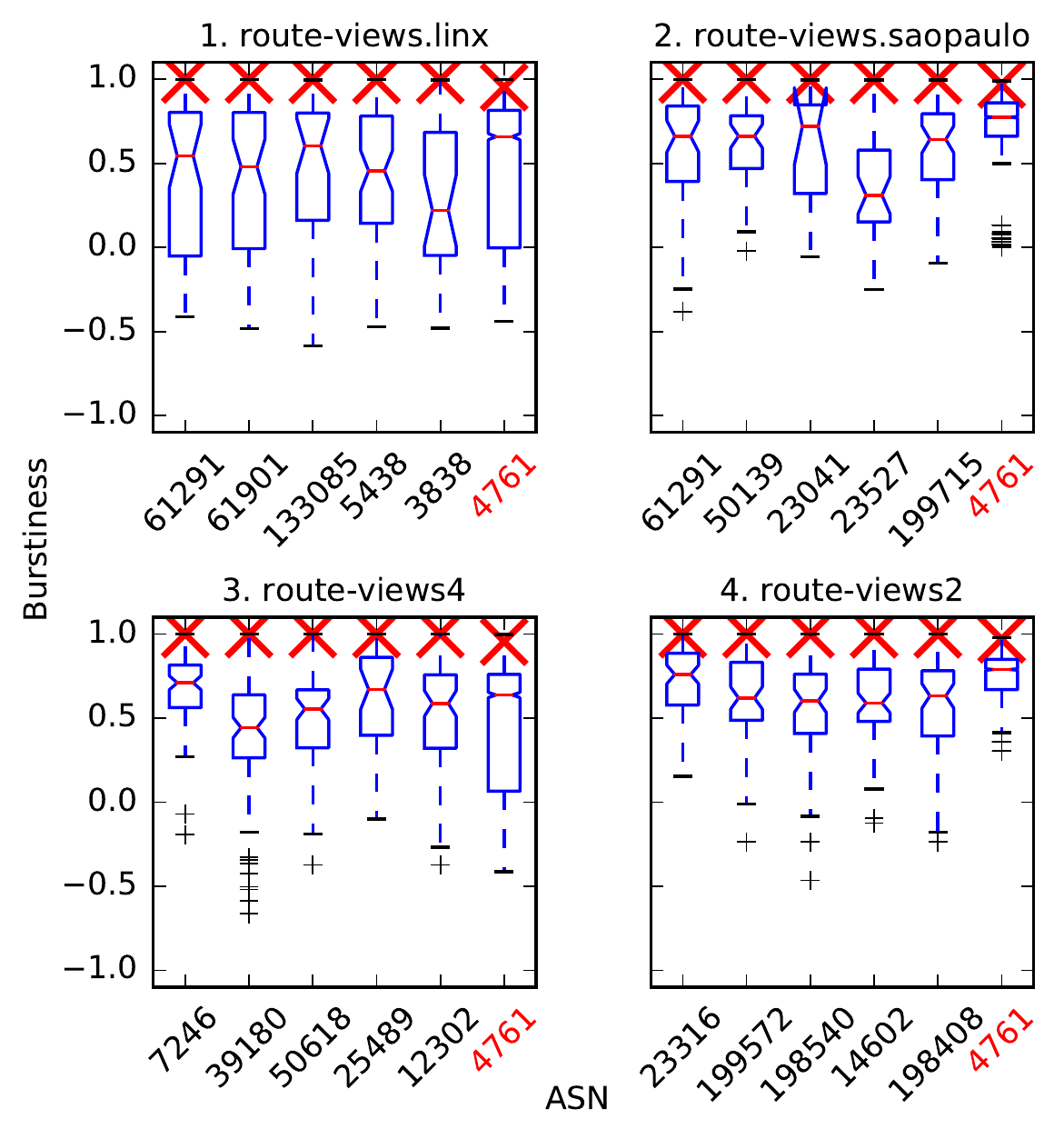} 
\caption{Monte Carlo test for burstiness. Last column corresponds to the observations of the AS responsible for the incident, i.e., AS4761. }
\label{fig-box-plot-null-case-indosat}
\end{figure}

Figure~\ref{fig-box-plot-null-case-indosat} shows notched box plots comparing the burstiness calculated for the baseline ASes and the AS involved in the incident (the last one) under the null hypothesis. Notched box plots have a contraction around the median whose height is statistically important. When notches of the boxes overlap, there is not a statistically significant difference between the medians. In this case, these plots illustrate that the burstiness of each of the ASes under study are not significantly different when there is no incident. The observation highlighted with the red X corresponds to the test statistic for the observations derived during the interval of the incident. As can be seen, for collectors receiving announcements from the AS involved in the incident, this observation lays outside the region of statistical indistinguishability. This suggests that the burstiness during the incident is statistically significant different, and it is unlikely that such values would be observed under random conditions. This argument reinforces the idea that the volume of announcements is a necessary but not sufficient feature for detection of BGP incidents (see Fig.~\ref{fig-joint-distribution-indosat}). High burstiness is a distinctive feature too.

\begin{figure}[!htbp]
\setlength\abovecaptionskip{-1.5\baselineskip}
\centering
\includegraphics[width=1.0\columnwidth]{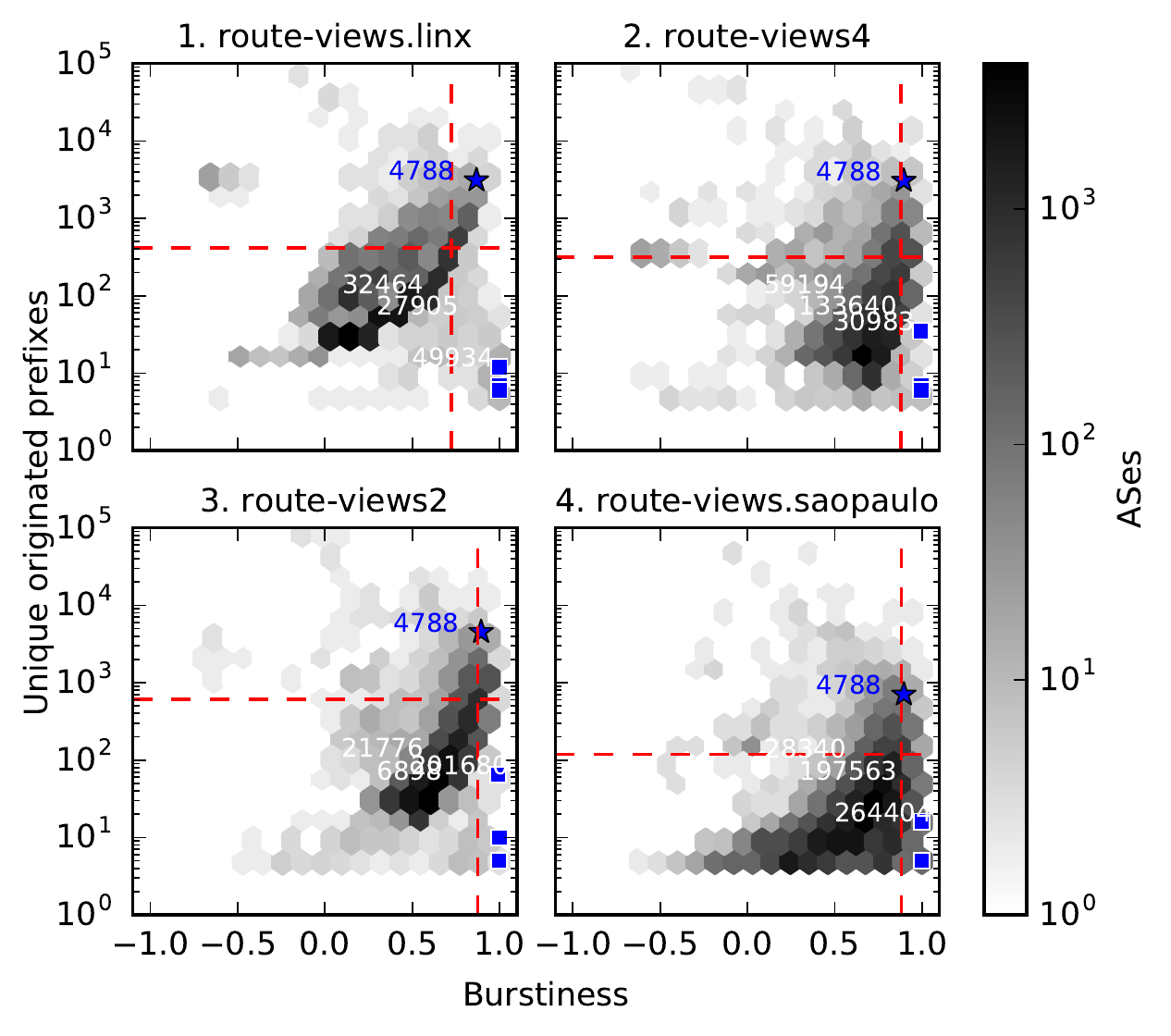} 
\caption{Joint distribution based on the total number of announcements and their burstiness during one day interval around the Telecom Malaysia incident.}
\label{fig-joint-distribution-malaysia}
\end{figure}

\begin{figure}[!htbp]
\setlength\abovecaptionskip{-1.5\baselineskip}
\centering
\includegraphics[width=1.0\columnwidth]{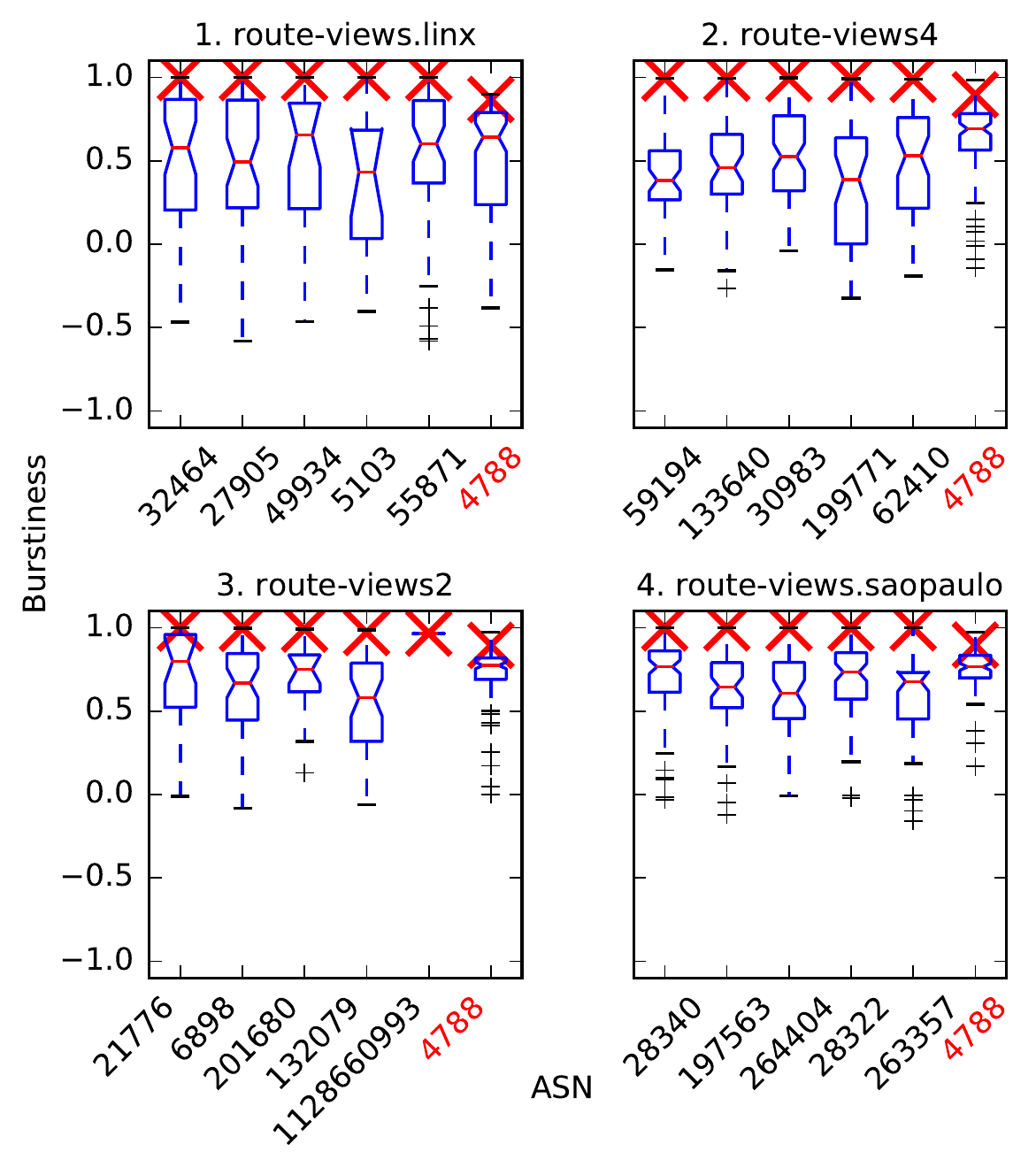} 
\caption{Monte Carlo test for burstiness. The last column corresponds to the observations of the AS responsible for the incident, i.e., AS4788.}
\label{fig-box-plot-null-case-malaysia}
\end{figure}

\noindent \textbf{Telecom Malaysia.} Figure~\ref{fig-joint-distribution-malaysia} shows that the AS involved in the incident has a distinct characterization in the distribution, i.e., AS4788. It has both high burstiness and number of announcements. Note that there are ASes that sent a high number of announcements and do not have high burstiness compared to Telecom Malaysia (those in the second quadrant), e.g., AS54169, AS28573, AS23752. These ASes were not involved with the incident nor are they neighbors of Telecom Malaysia. Conversely, the ASes in the fourth quadrant have higher burstiness but fewer announcements compared to Telecom Malaysia, e.g., AS134036, AS50710, that is consistently found among collectors. These ASes are not neighbors of Telecom Malaysia, and there is no evidence of malicious updates coming from them.  
Figure~\ref{fig-box-plot-null-case-malaysia} shows the distribution of burstiness computed over $100$ samples of random one-day intervals. This figure shows that the burstiness of the AS that was involved in the incident is statistically significantly larger when compared to its own normal behavior (e.g., baseline and null comparisons).

\noindent \textbf{Rostelecom.} We refer readers to~\ref{Inter-arrival time analysis: Rostelecom}. We found similar findings as the Indosat and Telecom Malaysia incidents. That means that for the Rostelecom incident, we found a statistically significant number of announcements as well as burstiness (see Fig.~\ref{fig-joint-distribution-russia}). We also found that the perpetrator's burstiness during the incident stands out with respect to itself (see Fig.~\ref{fig-box-plot-null-case-russia}).

\subsection{Anomaly detection} \label{Anomaly detection}

The main idea of our anomaly detection method relies on profiling the expected behavior of a signal and then detecting deviations from the expected pattern. To do so, we rely on the previous finding that revealed that there is a significant increase in the burstiness of announcements when an incident happens. We operationalize that insight by computing the time series $Q_{\alpha \to \beta}(t)$ for each incident, based on Eq.~\eqref{eq:Q computation}. We notice that analyzing the volume of announcements can be misleading, and adding the measure of burstiness has two advantages. First, a high volume of announcements may be caused by BGP session resets and other vendor specific  behaviors \cite{Wang:2002:GBP:Under:Stress}. Second, it enables detection of anomalies and decreases the number of candidates to be examined as potential anomalies (e.g., quadrant two in Figs.~\ref{fig-joint-distribution-indosat}, \ref{fig-joint-distribution-malaysia}).  

In the following analysis, the time series of $Q_{\alpha \to \beta}(t)$ is represented by the circles.\footnote{Each individual circle corresponds to the value of $Q_{\alpha \to \beta}(t)$. We did not bind them.} The solid line represents the EMA of $Q_{\alpha \to \beta}(t)$ for each arriving unique announcement message at time $t$. Each horizontal gray band represents one standard deviation from the moving average using the same window length. The darkness of the bands indicates the distance from the means based on Algorithm~\ref{alg-Algorithm-Event-Detection}. Observations that are more than two standard deviations away from the EMA are marked with stars. We finally compare the performance of the proposed method (based on EMA and LSTM) and the baseline of volume (see Figs.~\ref{fig-time-series-announcements-indosat},~\ref{fig-time-series-announcements-malaysia},~\ref{fig-time-series-announcements-russia}), denoted by ``P'' and ``V'' respectively, and report their results based on precision, recall, and F1 score (to take care of the unbalance nature of the dataset). We described the details about the LSTM in Section~\ref{Time series anomaly detection}. We used a detection resolution $m=3$ hours because it is the minimum length of a sustained burst of announcements according with the incidents that we studied. We emphasized in bold the result that achieves better performance with respect to a particular metric. The procedure to compute the performance was described in Section~\ref{Detection evaluation}. Here, we provide the results for the top four collectors based on the number of feeders. For the remaining collectors please refer to the Supplement, Section~\ref{Remaining collectors appendix}. 

\begin{figure}[!htbp]
\setlength\abovecaptionskip{-1.5\baselineskip}
\centering
\includegraphics[width=1.0\columnwidth]{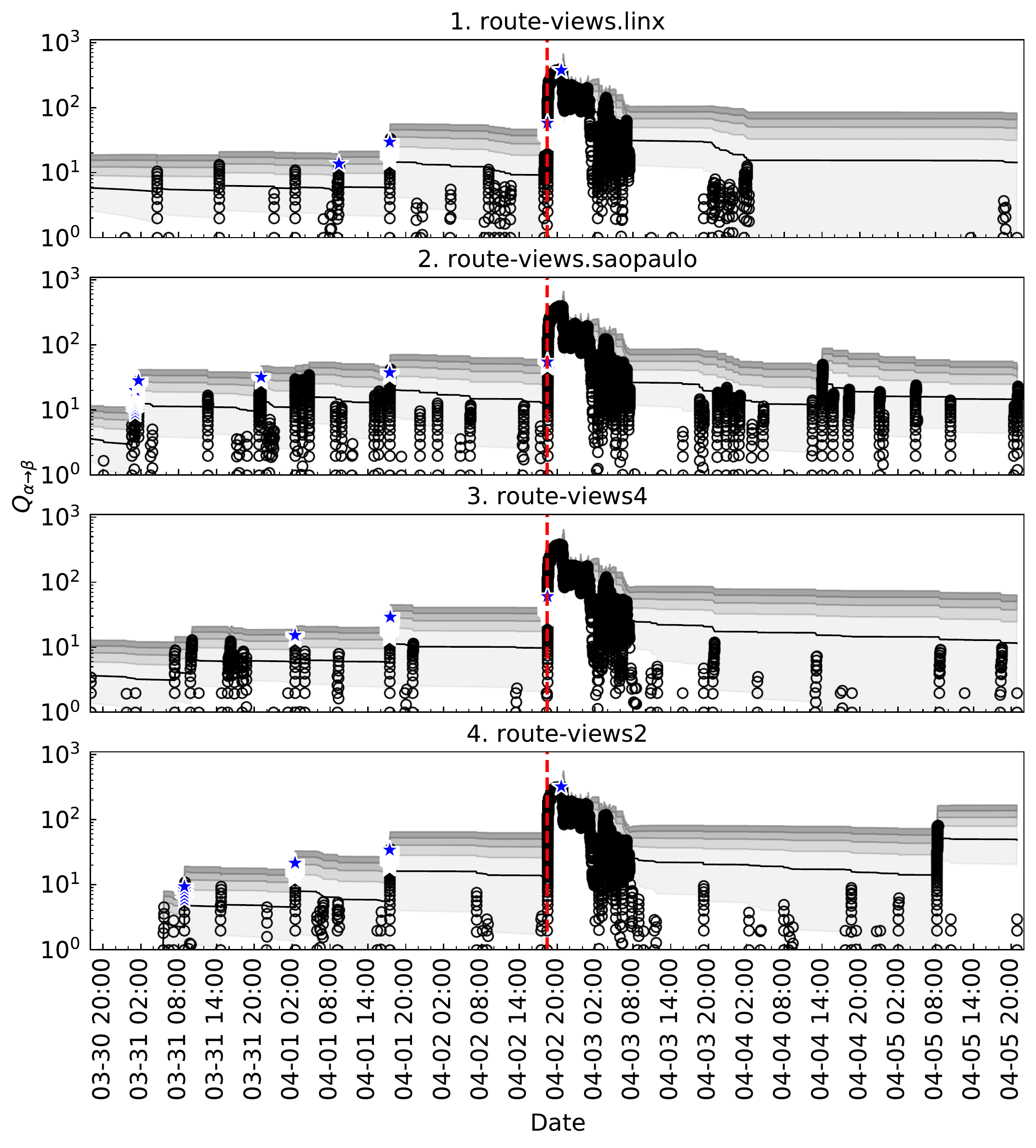} 
\caption{$Q_{4761 \to \beta}$ time series for the Indosat incident.}
\label{fig-Q-measure-Indonesia}
\end{figure}

\noindent \textbf{Indosat.} Figure~\ref{fig-Q-measure-Indonesia} shows that around when the event was reported, $Q_{4761 \to \beta}$ is more than two standard deviations away from the EMA. We observe the correlation between the significant observations and the ground truth. Note also that there are other outliers before the start of the incident. We do not have evidence of other anomalies occurring at these other specific times. Table~\ref{Table:Indonesia performance} shows the results of the performance comparison. Note that the proposed method achieves a lower number of false positives while still detecting the incident in the shown collectors. This is true for both detection methods based on EMA and LSTM. We observe that in general, the result of the anomaly detection based on EMA tends to outperform the one based on LSTM. This is in agreement with previous findings for forecasting time series \cite{Makridakis:2018:Stats:ML:Forecasting}. We discuss insights about anomalous announcements that were not reported by the ground and its impact on improving precision in the Supplement, Section~\ref{Performance results with augmented ground truth appendix}.

\begin{table}[htp!] 
\caption{Performance comparison for the Indosat incident.} 
\centering 
\small
\tabcolsep=0.01cm
\begin{tabularx}{0.49\textwidth}{| C | l | C | C | C | C | C | C |} 
\hline
& \multicolumn{1}{D}{\multirow{2}{*}{\textbf{Collector name}}} & \multicolumn{2}{|D}{\textbf{Precision}} & \multicolumn{2}{|D}{\textbf{Recall}} & \multicolumn{2}{|D|}{\textbf{F1 score}} \\
\cline{3-8} 
& & P & V & P & V & P & V \\
\hline
\hline
\parbox[t]{2mm}{\multirow{4}{*}{\rotatebox[origin=c]{90}{EMA}}} & route-views.linx & \textbf{25\%} & 6.7\% & 100\% & 100\% & \textbf{40\%} & 12.5\% \\ 
& route-views.saopaulo & \textbf{25\%} & 3.7\% & 100\% & 100\% & \textbf{40\%} & 7.1\%  \\ 
& route-views4 & \textbf{33.3\%} & 7.1\% & 100\% & 100\% & \textbf{50\%} & 13.3\%  \\ 
& route-views2 & 0\% & \textbf{5.9\%} & 0\% & \textbf{100\%} & 0\% & \textbf{11.1\%}  \\ 
\hline
\parbox[t]{2mm}{\multirow{4}{*}{\rotatebox[origin=c]{90}{LSTM}}} & route-views.linx & \textbf{7.1\%} & 6.7\% & 100\% & 100\% & \textbf{13.3\%} & 12.5\% \\ 
& route-views.saopaulo & \textbf{5.6\%} & 3.7\% & 100\% & 100\% & \textbf{10.5\%} & 7.1\%  \\ 
& route-views4 & 7.1\% & 7.1\% & 100\% & 100\% & 13.3\% & 13.3\%  \\ 
& route-views2 & \textbf{10\%} & 5.9\% & 100\% & 100\% & \textbf{18.2\%} & 11.1\%  \\ 
\hline
\end{tabularx} 
\label{Table:Indonesia performance} 
\end{table}

\begin{figure}[!htbp]
\setlength\abovecaptionskip{-1.5\baselineskip}
\centering
\includegraphics[width=1.0\columnwidth]{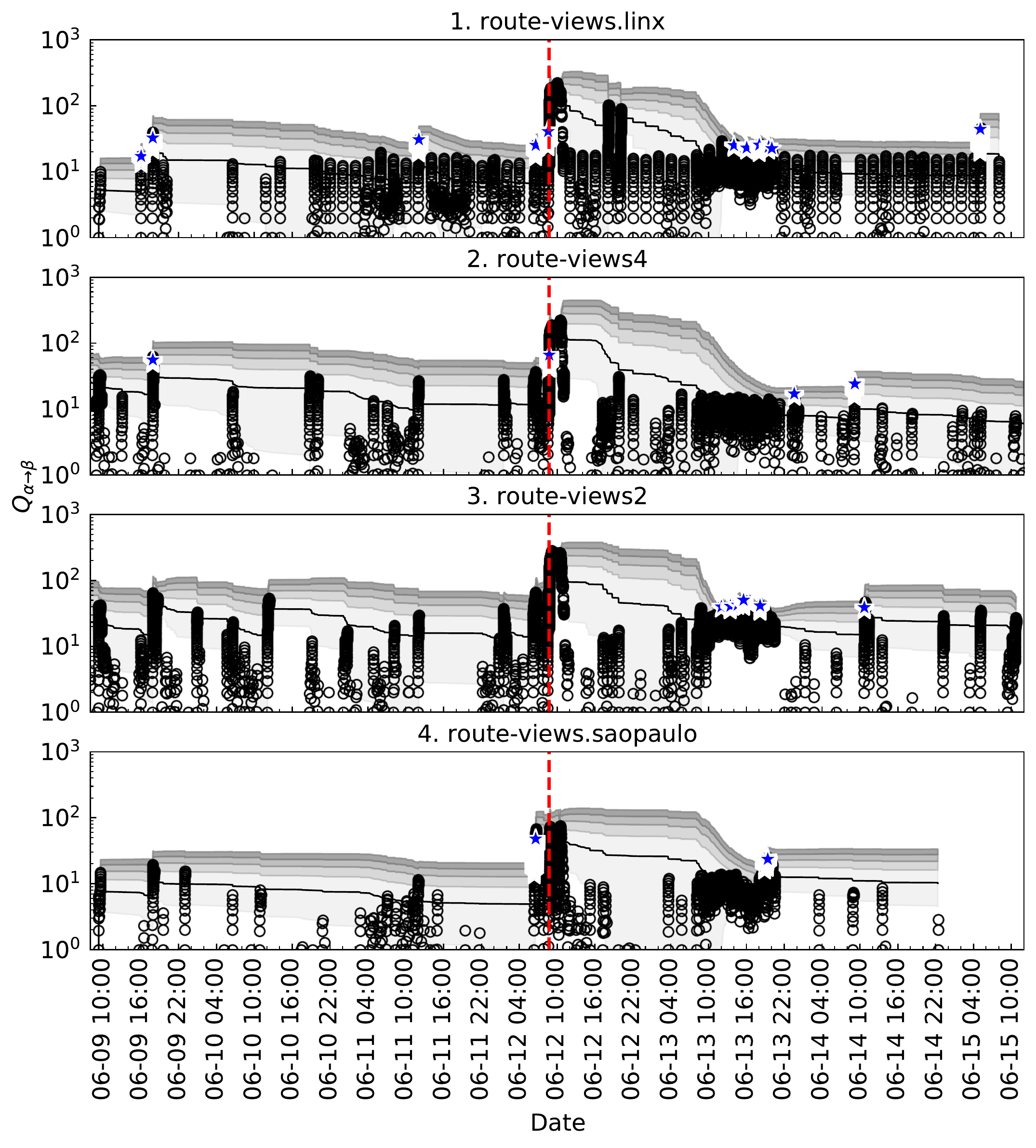} 
\caption{$Q_{4788 \to \beta}$ time series for the Telecom Malaysia incident.}
\label{fig-Q-measure-Malaysia}
\end{figure}

\noindent \textbf{Telecom Malaysia.} Figure~\ref{fig-Q-measure-Malaysia} shows the time series of $Q_{\alpha \to \beta}$. The value of $Q_{4788 \to \beta}$ is more than two standard when the incident was reported by BGPMon. Note also that route-views2 reports no outliers correlated with the ground truth. This means that the perceived burstiness is not as high as for the other collectors (see Fig.~\ref{fig-joint-distribution-malaysia}). Table~\ref{Table:Malaysia performance} shows the results of the performance comparison. Overall, the proposed method outperforms the baseline of volume for both EMA and LSTM helping to reduce the number of false negatives. 

\begin{table}[htp!] 
\caption{Performance comparison for the Telecom Malaysia incident.} 
\centering 
\small
\tabcolsep=0.01cm
\begin{tabularx}{0.49\textwidth}{| C | l | C | C | C | C | C | C |} 
\hline
& \multicolumn{1}{|D}{\multirow{2}{*}{\textbf{Collector name}}} & \multicolumn{2}{|D}{\textbf{Precision}} & \multicolumn{2}{|D}{\textbf{Recall}} & \multicolumn{2}{|D|}{\textbf{F1 score}} \\
\cline{3-8} 
& & P & V & P & V & P & V \\
\hline
\hline
\parbox[t]{2mm}{\multirow{4}{*}{\rotatebox[origin=c]{90}{EMA}}} & route-views.linx & \textbf{14.3\%} & 2.5\% & 100\% & 100\% & \textbf{25\%} & 4.9\% \\ 
& route-views4 & \textbf{20\%} & 4.8\% & 100\% & 100\% & \textbf{33.3\%} & 9.1\%  \\ 
& route-views2 & 0\% & \textbf{4.2\%} & 0\% & \textbf{100\%} & 0\% & \textbf{7.9\%}  \\ 
& route-views.saopaulo & \textbf{33.3\%} & 9.1\% & 100\% & 100\% & \textbf{50\%} & 16.7\%  \\ 
\hline
\parbox[t]{2mm}{\multirow{4}{*}{\rotatebox[origin=c]{90}{LSTM}}} & route-views.linx & \textbf{5.3\%} & 2.5\% & 100\% & 100\% & \textbf{10\%} & 4.9\% \\ 
& route-views4 & \textbf{25\%} & 4.8\% & 100\% & 100\% & \textbf{40\%} & 9.1\%  \\ 
& route-views2 & 0\% & \textbf{4.2\%} & 0\% & \textbf{100\%} & 0\% & \textbf{7.9\%}  \\ 
& route-views.saopaulo & \textbf{10\%} & 9.1\% & 100\% & 100\% & \textbf{18.2\%} & 16.7\%  \\ 
\hline
\end{tabularx} 
\label{Table:Malaysia performance} 
\end{table}

\begin{figure}[!htbp]
\setlength\abovecaptionskip{-1.5\baselineskip}
\centering
\includegraphics[width=1.0\columnwidth]{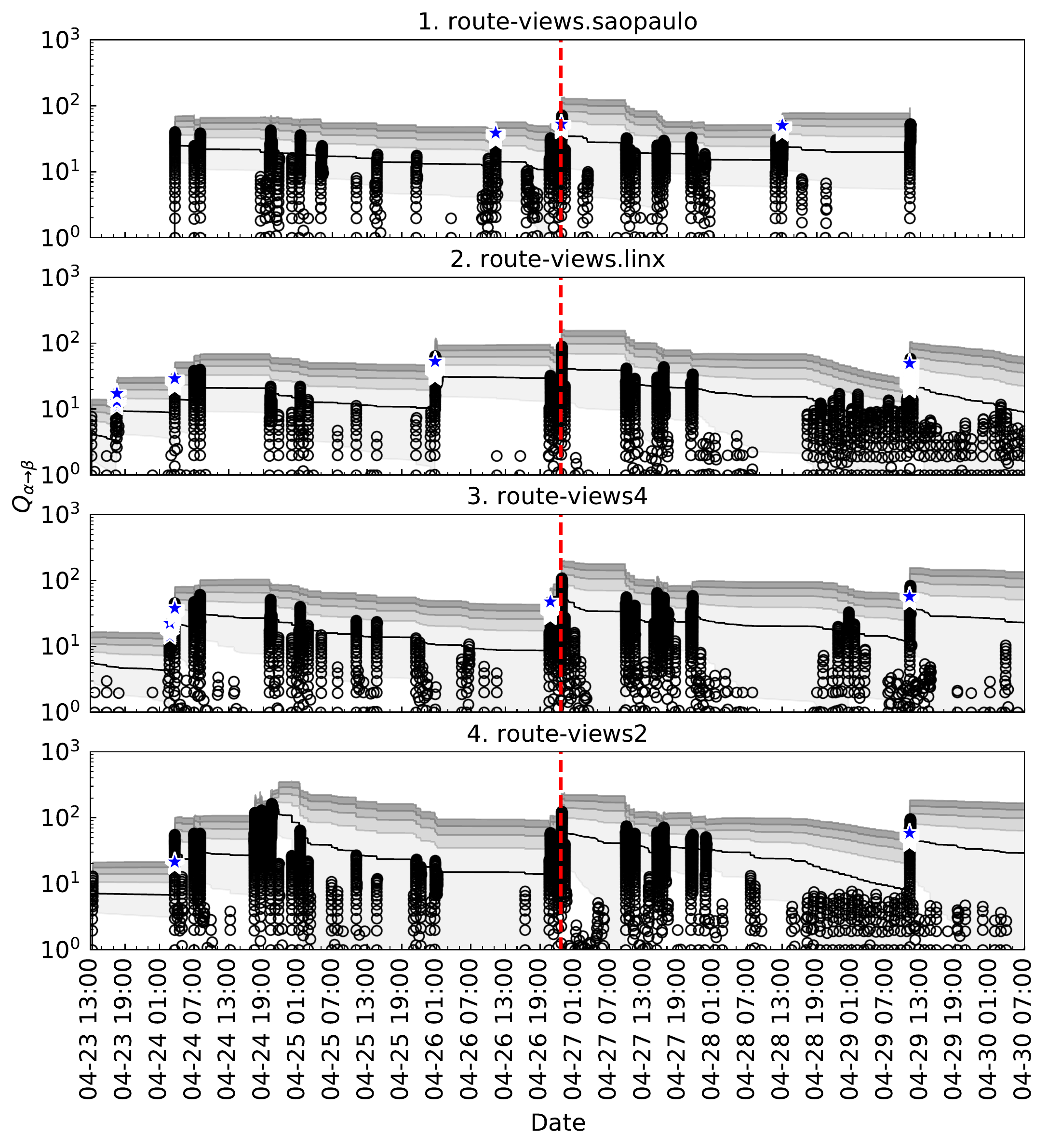} 
\caption{$Q_{12389 \to \beta}$ time series for the Rostelecom incident. }
\label{fig-Q-measure-Russia}
\end{figure}

\noindent \textbf{Rostelecom.} Figure~\ref{fig-Q-measure-Russia} shows that route-views.saopaulo can detect the incident and is correlated with the ground truth. This is consistent with Fig.~\ref{fig-joint-distribution-russia} that shows that the perpetrator is placed in the first quadrant (i.e., significant burstiness and number of announcements). The proposed method is able to detect the incident as well in route-views4 and but does not see it in route-views.linx nor route-views2. Table~\ref{Table:Russia performance} shows the results of the performance comparison for this incident. Overall, the proposed method outperforms the baseline of volume while still detecting the incident. EMA based prediction tends to produce better results than the LSTM overall.

\begin{table}[htp!] 
\caption{Performance comparison for the Rostelecom incident.} 
\centering 
\small
\tabcolsep=0.01cm
\begin{tabularx}{0.49\textwidth}{| C | l | C | C | C | C | C | C |} 
\hline
& \multicolumn{1}{D}{\multirow{2}{*}{\textbf{Collector name}}} & \multicolumn{2}{|D}{\textbf{Precision}} & \multicolumn{2}{|D}{\textbf{Recall}} & \multicolumn{2}{|D|}{\textbf{F1 score}} \\
\cline{3-8} 
& & P & V & P & V & P & V \\
\hline
\hline
\parbox[t]{2mm}{\multirow{4}{*}{\rotatebox[origin=c]{90}{EMA}}} & route-views.saopaulo & \textbf{33.3\%} & 0\% & \textbf{100\%} & 0\% & \textbf{50\%} & 0\% \\ 
& route-views.linx & 0\% & 0\% & 0\% & 0\% & 0\% & 0\%  \\
& route-views4 & \textbf{33.3\%} & 0\% & \textbf{100\%} & 0\% & \textbf{50\%} & 3.5\%  \\
& route-views2 & 0\% & 0\% & 0\% & 0\% & 0\% & 0\%  \\
\hline
\parbox[t]{2mm}{\multirow{4}{*}{\rotatebox[origin=c]{90}{LSTM}}} & route-views.saopaulo & \textbf{14.3\%} & 0\% & \textbf{100\%} & 0\% & \textbf{25\%} & 0\% \\ 
& route-views.linx & 0\% & 0\% & 0\% & 0\% & 0\% & 0\%  \\
& route-views4 & 0\% & 0\% & 0\% & 0\% & 0\% & 0\%  \\
& route-views2 & \textbf{25\%} & 0\% & \textbf{100\%} & 0\% & \textbf{40\%} & 0\%  \\
\hline
\end{tabularx} 
\label{Table:Russia performance} 
\end{table}

\section{Discussion} \label{Discussion}

Routing anomalies caused by both misconfigurations and malicious intent have tested the resilience of Internet core protocols \cite{Moriano:2017:Macro:BGP}. Here, we propose an anomaly detection method and show that it would identify different BGP incidents in agreement with manually verified ground truth. 
To do so, we analyze inter-arrival times of BGP announcements leveraging the RouteViews collector infrastructure. We found that the burstiness, along with the volume of announcements, has the potential to provide warnings of routing anomalies when they are evident using traditional control-plane and data-plane approaches. 

To validate the effectiveness of the proposed method, we conducted analysis for six cases of routing anomalies (see Section~\ref{Ground truth} for more details). We have evaluated the statistical significance of announcement burstiness, before, during, and after the events. We found that the perpetrators of the incidents have statistically significant bursty patterns that are visible from some collectors. We analyze the same features under the null case (of no incidents) and corroborate that the bursty behavior is characteristic of announcements  sent prior the detection of the incidents. By relying on this key observation, we propose an algorithm to identify when there is an incipient anomalous incident. We made the data and scripts used in this research for reproducibility purposes. 

The proposed method would be effective against hijacks, route leaks, and incidents leading to traffic interception. Having noted the potential for our approach, we are also aware of some limitations of our proposed work. 

\noindent \textbf{Real-time data availability:} Our analysis is based on BGP announcements received by RouteViews collectors. Only a subset of these collectors support real-time monitoring through BGPmon.\footnote{Here BGPmon refers to the free monitoring service develop by Colorado State University available at \href{https://www.bgpmon.io/}{https://www.bgpmon.io/}} The RouteViews data used in our analysis relies on BGPStream, which has an access delay of approximately 20 min \cite{Sermpezis:2018:Artemis}. One option for further research is to run these experiments with a reduced number of current real-time RouteViews collectors through BGPmon. In addition, RIPE RIS provides an API to access real-time BGP updates for a limited number of collectors. Through sharing our scripts, we hope that individual collectors could implement this approach and report the results in the future.

\noindent \textbf{Feeder contribution:} Our method treats each router contribution as equivalent. However, they vary significantly in terms of IP space coverage as shown in previous research \cite{Gregori:2012:BGP:Route:Collector, Gregori:2015:AS-level:Incompleteness, Testart:2020:RPKI:Filtering}. This has an effect on the view that each collector has and the detection of the incidents.

\noindent \textbf{Focus on detection but not mitigation:} We propose an anomaly detection method that allows the identification of BGP incidents. To do so, the effectiveness of our proof-of-concept is evaluated based on its ability to detect incidents with respect to manually verified ground truth metadata. Yet we do not discuss mitigation strategies once the events are detected, e.g., prefix deaggregation \cite{Lutu:2012:Economics:Prefix:Deaggregation}. Of course, these mitigation strategies can be implemented on top of our proposed method to avoid wide diffusion of route misinformation. 

 \noindent \textbf{Ground truth:} We compute the performance of the detection task based on the ground truth described in Section~\ref{Ground truth}. We, however, noticed that beyond the reported ground truth, there are maybe other invalid announcements before or after the reported ground truth. Precision and F1 score metrics are affected because we restricted our ground truth for performance comparison. By augmenting the ground truth with the announcement of fake prefixes, we show that precision can be improved at the expense of recall. Further analysis on this is discussed in the Supplement, Section~\ref{Performance results with augmented ground truth appendix}.

\section{Conclusion} \label{Conclusion}

We illustrate the efficacy of leveraging RouteViews collectors' infrastructure to identify anomalous routing incidents through the analysis of inter-arrival times of announcements. As a complement to current anomaly identification approaches, we have demonstrated a proof-of-concept that identifies real hijack incidents when these were detected in practice by leveraging the current RouteViews collectors' infrastructure. We have characterized six different incidents, including, large-scale and for traffic interception purposes from a different perspective, one derived by analyzing the patterns of burstiness of BGP announcements. The method detailed in this paper relies on the fact that large-scale disruption events produces groups of BGP announcements of relatively high frequency followed by periods of relatively infrequent events, which can be measured as burstiness. Relying on this observation, we describe a detection method that is able to indicate, from a collector point of view, when an incident is incipient.

Additional future work includes examining the effectiveness of the proposed method with real-time BGP updates from different collector projects. BGPmon provides real-time BGP feeds from several feeders as well as some collectors in the RIPE RIS project. The approach in this paper can be also tested with a protocol specifically designed for monitoring purposes, such as the OpenBMP protocol \cite{Scudder:2016:OpenBMP}. An implementation of a prototype for anomaly detection based on the principles of this paper seems feasible with the availability of real-time data from different projects available in BGPStream.

\section*{Acknowledgments}

This research has been supported in part by NSF CNS 1565375 and Cisco Research 591000. This work was carried out in part at Oak Ridge National Laboratory, managed by UT-Battelle, LLC for the U.S. Department of Energy under Contract No. DE-AC05-00OR22725. Pablo Moriano thanks Kalyan Perumalla and Steve Rich for their guidance, and Claudia Castro for her assistance in designing Figure~\ref{fig-lstm-cell}.

\appendix

\section{Inter-arrival time analysis: Rostelecom} \label{Inter-arrival time analysis: Rostelecom}

\begin{figure}[!htbp]
\setlength\abovecaptionskip{-1.5\baselineskip}
\centering
\includegraphics[width=1.0\columnwidth]{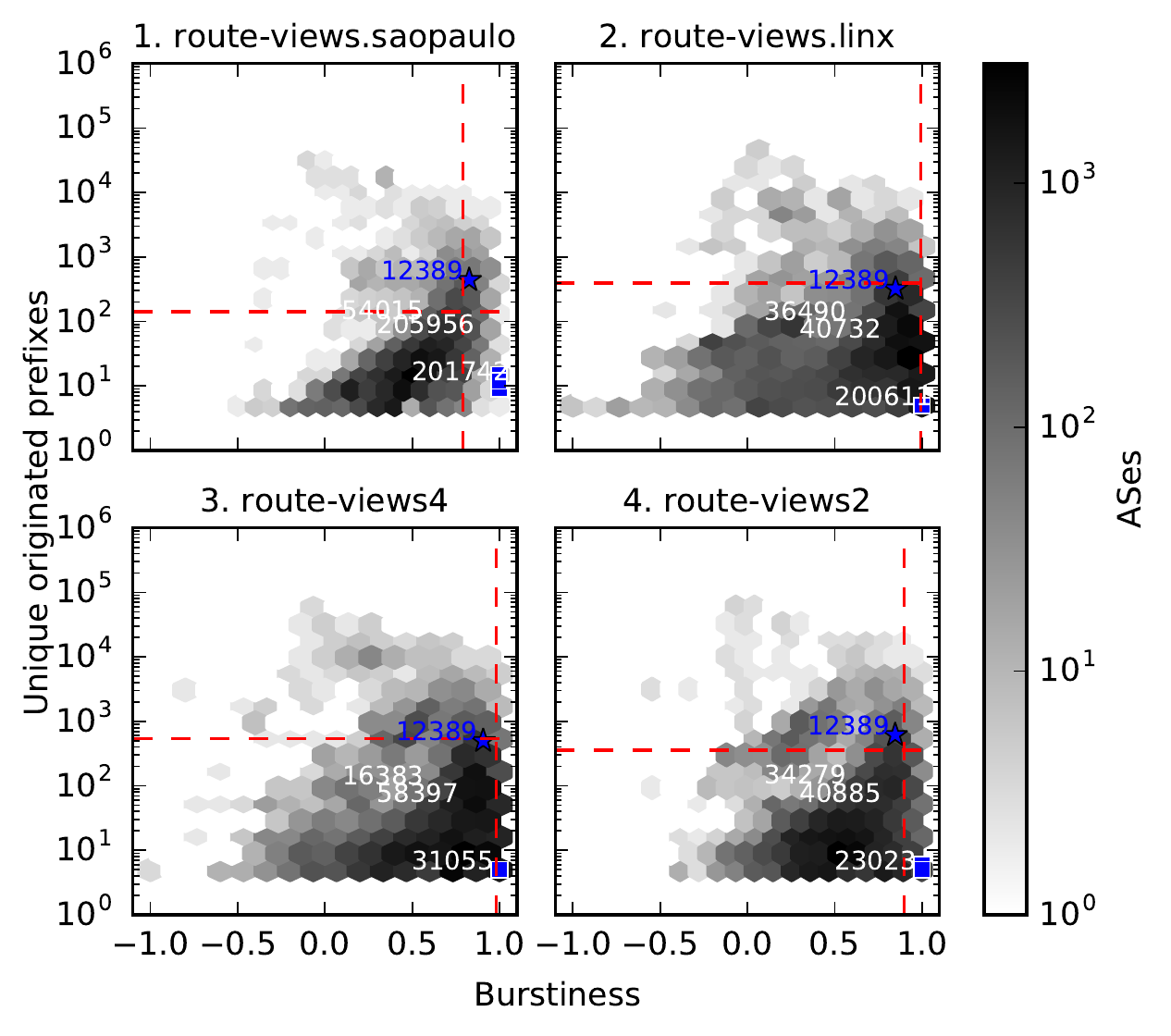} 
\caption{Joint distribution based on the total number of announcements and their burstiness during the one day interval around the Rostelecom incident.}
\label{fig-joint-distribution-russia}
\end{figure}

Figure~\ref{fig-joint-distribution-russia} shows the placement of AS12389 in the joint distribution. Overall for this incident, the perpetrator tends to have sent a high number of announcements and a significant bustiness for route-views.saopaulo. Other ASes with a significant number of announcements include AS15133, AS3203, and AS29049. These collectors are not neighbors of AS12389. 

\begin{figure}[!htbp]
\setlength\abovecaptionskip{-1.5\baselineskip}
\centering
\includegraphics[width=1.0\columnwidth]{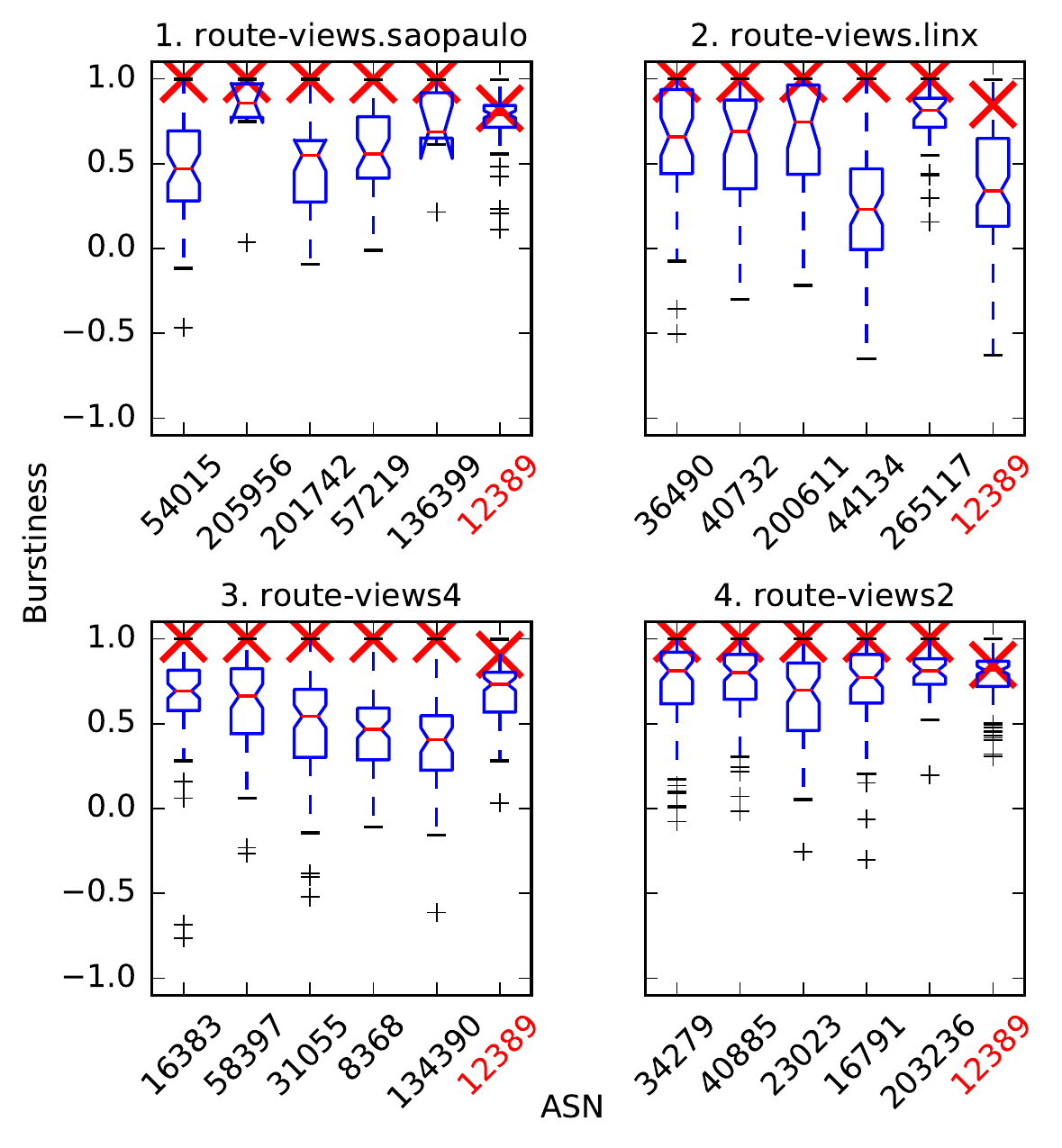} 
\caption{Monte Carlo test for burstiness. The last column corresponds to the observations of the AS responsible for the incident, i.e., 12389.}
\label{fig-box-plot-null-case-russia}
\end{figure}

Figure~\ref{fig-box-plot-null-case-russia} shows the bustiness of the perpetrator during the incident and a comparison with itself and the top five burstiest ASes. Overall, the bustiness of the perpetrator is not as high as the top ASes during the incident but it stands out with respect to itself. 


\bibliographystyle{elsarticle-num}
\bibliography{/Users/pmoriano/Dropbox/Bib/Bibliography} 




%
\end{document}